\documentclass[journal]{IEEEtran}
\usepackage{cite}
\usepackage{graphicx}
\usepackage{subfigure}

\usepackage{xcolor,soul,framed} %,caption

\colorlet{shadecolor}{yellow}
\DeclareGraphicsExtensions{.pdf,.jpeg,.png,.jpg}

\usepackage[cmex10]{amsmath}
\usepackage{array}
\usepackage{mdwmath}
\usepackage{mdwtab}
\usepackage{eqparbox}
\usepackage{url}

\begin{document}
\bstctlcite{IEEEexample:BSTcontrol}
    \title{Micro-Vibration Modes Reconstruction Based on Micro-Doppler Coincidence Imaging }
  \author{Shuang~Liu,~Chenjin~Deng,~Chaoran~Wang,~Zunwang~Bo,~Shensheng~Han,~Zihuai~Lin,~\IEEEmembership{Senior Member,~IEEE}}
      % <-this % stops a space

%\markboth{IEEE TRANSACTIONS ON }

% ====================================================================
\maketitle

% === ABSTRACT ====================================================================
% =================================================================================
\begin{abstract}
%\boldmath
Micro-vibration, a ubiquitous nature phenomenon, can be seen as a characteristic feature on the objects, these vibrations always have tiny amplitudes which are much less than the wavelengths of the sensing systems, thus these motions information can only be reflected in the phase item of echo. Normally the conventional radar system can detect these micro vibrations through the time frequency analyzing, but these vibration characteristics can only be reflected by time-frequency spectrum, the spatial distribution of these micro vibrations can not be reconstructed precisely. %And ghost imaging(GI), which is a novel imaging method originated quantum and optical areas, can reconstruct an unknown images by computational methods, but most of GI schemes reported in literature or in prototype only show the method to acquire reflection intensity images, the information within phase item of target which contains micro vibration characteristics gets lost.
Ghost imaging (GI), a novel imaging method also known as Coincidence Imaging that originated in the quantum and optical fields, can reconstruct unknown images using computational methods. 
To reconstruct the spatial distribution of micro vibrations, this paper proposes a new method based on a coincidence imaging system. A detailed model of target micro-vibration is created first, taking into account two categories: discrete and continuous targets. We use the first-order field correlation feature to obtain objective different micro vibration distribution based on the complex target models and time-frequency analysis in this work.
%Firstly, the detailed model of target micro vibration is set, considering the resolution of different remote sensing scenarios, the objective model can be divided into two cases to be discussed: discrete target and continuous target. Using the feature of first-order field correlation to obtain the complex images of the target plane. Based on the results of echo time-frequency analyze and target models, several methods are proposed to obtain objective high dimensional information. It provides system support for further target recognition based on high dimensional information.
\end{abstract}

% === KEYWORDS ====================================================================

% \fi
\IEEEpeerreviewmaketitle

% ====================================================================
% ====================================================================
% ====================================================================

% === I. INTRODUCTION =============================================================
% =================================================================================
\section{Introduction}

\IEEEPARstart{M}{icro}-vibration widely exists on the target surfaces as a motion characteristics, and there are many researches aiming at these micro-vibrations detecting\cite{MV1,MV2}. These vibration detecting methods can be applied in many scenarios like vital signs detection \cite{MVD1,MVD2,IREALCARE5}, large structure vibration monitoring \cite{VD2} or environment monitoring based on IoT \cite{leng2020implementation,IoT_FD,RF_energy1}. Normally, the vibrating objects contains two types of information, temporal change and spatial distribution. Existing single temporal information sensing systems can reconstruct the time-frequency variation characteristics of vibrating objects in the field of view by analyzing micro-Doppler signatures\cite{VD3}, whereas the spatial information can not be reconstructed efficiently. Even though there exist detection schemes of spatial vibration information.
%Even though there are detection schemes of spatial vibration information are proposed%此处加引%， 
%it is limited by imaging efficiency.\\

%The ghost imaging(GI), which is also known as correlation imaging, is a non-local imaging method, which reconstruct an unknown image by computing the correlation function between transmitting light, which passes through an object and be collected by a point detector with no spatial resolution (the energy in the aperture is collected to a point by a lens, also known as bucket detector), and the reference radiation fields whose space and time varying modes are formed specifically\cite{OGI1,OGI2,OGI3,OGI4,OGI5,OGI6}.
Ghost imaging (GI), also known as correlation imaging, is a non-local imaging method that reconstructs an unknown image by computing the correlation function between speckle field that passes through an object and one-dimensional signal received by a bucket detector which can collect the energy in the aperture. The spatial and temporal coding patterns modulated on the reference radiation fields are formed specifically\cite{OGI1,OGI2,OGI3,OGI4,OGI5,OGI6}. In principle, correlation imaging is to form an irradiated object with spatial fluctuation by modulating the radiation source, and reconstruct the image of the object by correlating it with the intensity of the received signal of a bucket detector. This novel imaging technique overcomes the antenna aperture-induced limitation in imaging resolution and offers the benefits of forward-looking, staring, and fast-shooting imaging, which may be employed in unmanned systems for monitoring and observation of crucial locations.
%This novel imaging algorithm break through the limitation of imaging resolution caused by antenna aperture, and has the advantages of forward-looking, staring and fast-shooting imaging which can be used in surveillance and observation for critical region on unmanned systems. 
The GI algorithm has been verified in different waveband systems, such as X-ray\cite{Xray1,Xray2,Xray3}, microwave\cite{MGI3,ZH1,ZH2,ZH3,GI_ICASSP,GI_ECC,GFKD}, terahertz\cite{Thz1,Thz2} and optical waveband\cite{OP1,OP2,OP3}.\\

In the present stage, it is easier for the optical system to detect the fluctuation of the light intensity rather than the complex light field, thus the light intensity detected by the bucket detector can be used for second-order correlation which is GI namely\cite{CGI1,CGI2,CGI3}. And it is possible to control and detect the form of a complex field for a microwave sensing system. Coding the transmitting microwave field and decoding the echo signal can obtain the target information, this imaging algorithm using the first-order field correlation is named as Coincidence Imaging.

\begin{figure}[htbp]
\centering
\includegraphics[width=1\columnwidth,height=0.55\linewidth]{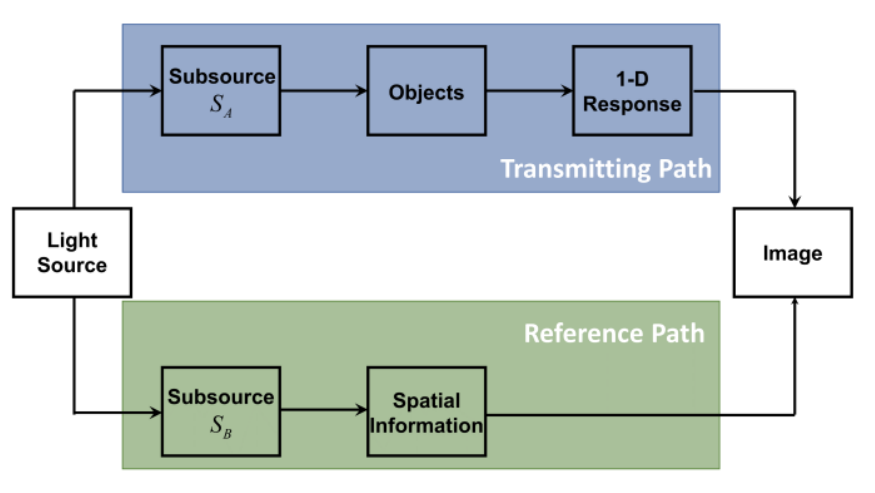}
\caption{The optical system structure of ghost imaging scheme. The light source is split into two identical subsources $S_A$ and $S_B$, the former is used to illuminate the object to be imaged and the latter is used as a reference. After the response from the object is received by a single-pixel detector, the final reconstructed image is obtained with the assistance from the reference spatial fluctuation information collected by a charge-coupled device(CCD) }
\label{fig：1}
\end{figure}
Most existing optical systems use the second-order correlation of light fields to reconstruct the object reflectivity image, in these scenarios, targets are modeled as real valued reflectivity in the field of view. The first order correlation method is proposed to realize phase imaging for static complex valued object planes. Considering that the vibration features of the object surface and the detection radiation field produce space-time modulation in essence, it is possible to extract vibrating information from the phase changes of the echo.
%For static complex valued object plane, the first order correlation method is proposed to realize phase imaging\cite{GWL}, and considering that the vibration features of the object surface and the detection radiation field produce space-time modulation in essence, so that vibrating information of object could be extracted from the phase changes of the echo.
However, considering that micro-vibration is a stale time-varying object characteristic, static phase recovery imaging can not satisfy the detection requirements of object vibrating features. There are some attempts\cite{PL1} to use the spatial correlation characteristic of ghost imaging to realize the spatial-temporal perception of vibrating, it has verified the feasibility of this theory. Considering the micro-vibration can be seen as a temporal-spatial modulation towards the radiation field on the surface of the target plane in essence, we model the vibration target as a dynamic complex reflectivity distribution in the field of view on the based of echo signal time-frequency analysis. The first-order field correlation imaging is intended to solve simultaneous spatial-temporal imaging of vibrating objects, it is crucial for the fulfillment of GI theory
%Micro-vibration ghost imaging is expected to solve the spatial-temporal imaging simultaneously of vibrating objects and it is of great significance for the completion of GI theory.
and this method can be applied to the other wavebands systems.\\

The outline of the paper is given below. Section II  introduces the first-order field correlation imaging theory and system architecture and Section III demonstrates the interaction model between micro-vibration object and radiation field. And the target characteristic models under two characteristics constraints are established in Section IV. Consequently, Section V presents micro-vibration reconstruction algorithms under the two different object models and their corresponding numerical simulations results are demonstrated as well.

\section{Static Target Phase Reconstruct in Coincidence Imaging Theory}
\begin{figure}[htpb]
\centering
\includegraphics[width=1\columnwidth,height=0.65\linewidth]{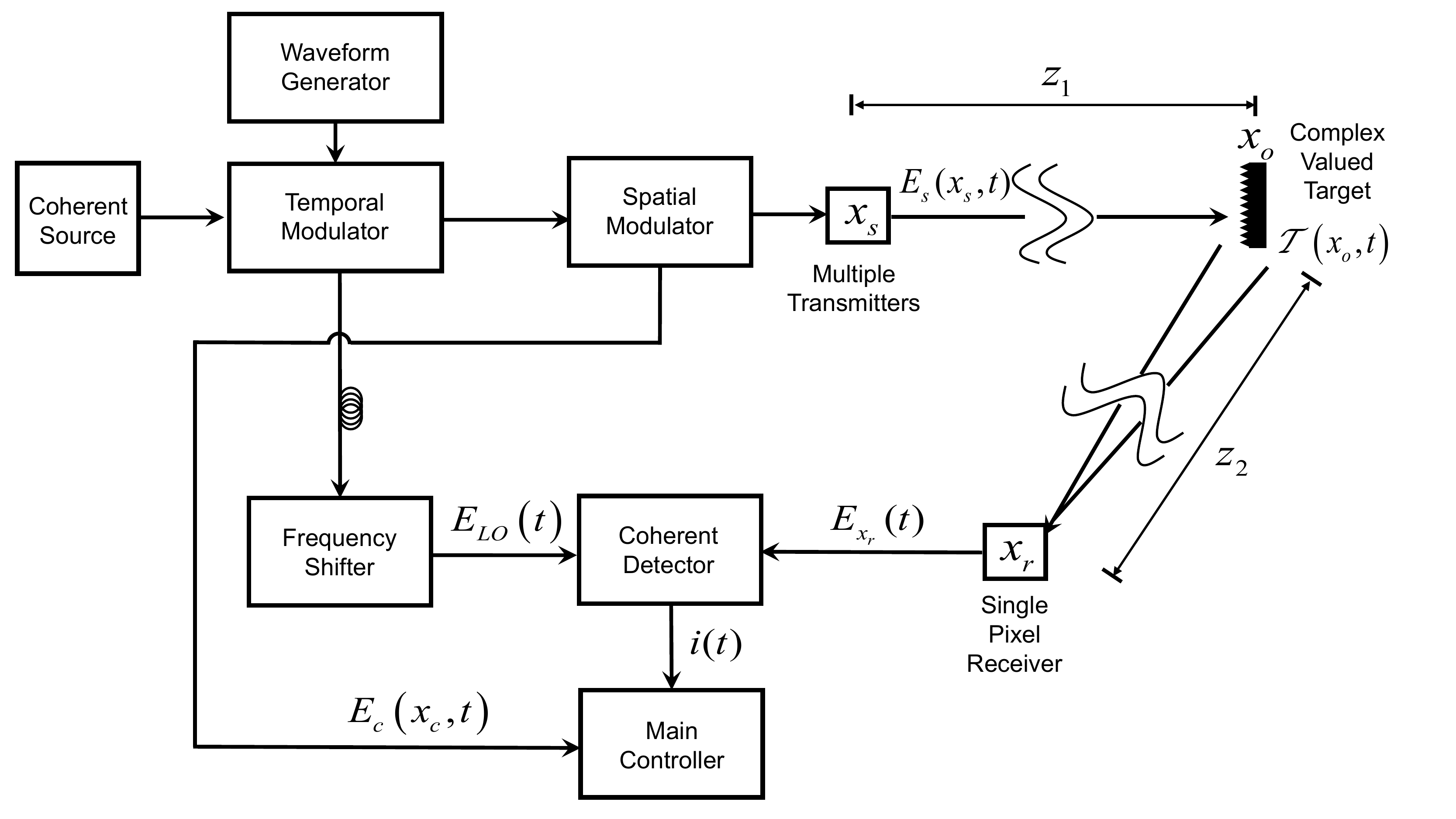}
\caption{Optical structure of the first-order field correlation systems, the target model is set as a complex valued target due to the micro-vibration.}
\label{fig：1}
\end{figure}
A novel micro-vibration mode detecting algorithm through the first-order field correlation is proposed in this section. Fig. 2 depicts the systematic diagram of a first-order field correlation radiation field imaging system.
%Set the optical system as an example, the systematic diagram of the first-order field correlation radiation field imaging system is shown in Fig.2.
The positions of the radiation fields generated in the whole imaging procession are annotated in Fig. 2, and their physical meanings are shown below, 
\begin{enumerate}
\item $E_s(x_s,t)$ is the radiation field with spatial distribution characteristics, which is spatial modulated by the prefabricated speckle patterns coding from a temporal modulated coherence radiation field source. $E_s(x_s,t)$ are transmitted by transmitters and stands for the radiation field on the transmitters plane, where $x_s$ stands for the the spatial coordinates of the transmitters plane
\item The radiation field reaching the target plane $x_o$ is denoted as $E_o(x_o,t)$ after the propagation at distance $Z_1$, it is worth noting that $E_o(x_o,t)$ has no interference with target characteristic at the moment.
\item $E_{x_r}(t)$ is the radiation field on the surface of receiver plane after the propagation at distance $Z_2$, and the propagation at $Z_1$ and $Z_2$  is denoted as transmitting path. Considering that single pixel receiver only receives a single point of coherent field, the receiving radiation field is a one-dimensional complex signal.
\item  $E_c(x_c,t)$ is the radiation field which can be calculated based on the system configuration and detecting information in an actual engineering system. The effect of reference field is to eliminate the phase interference of system and propagation by coherence in essence. $E_c(x_c,t)$ is denoted as reference path.
\item $E_{LO}(t)$ is the local oscillator($LO$) radiation field with no spatial characteristics output from frequency shifter. 
\item $i(t)$ is one-dimensional electronic signal output from coherent detector and this electronic signal is obtained by the coherence of the $LO$ radiation field $E_{LO}(t)$ and the received radiation field $E_{x_r}(t)$. $i(t)$ can have first-order field correlation with the reference radiation field $E_{c}(x_c,t)$.
\end{enumerate}
%A temporal modulated laser source can be spatial modulated by the prefabricated speckle pattern coding. The radiation field $E_s(x_s,t)$, where $x_s$ stands for the the spatial coordinates of the transmitters plane, with spatial distribution characteristics are transmitted by transmitters. After propagation at distance $z_1$, the radiation field reaches the target plane $x_o$ is denoted as $E_o(x_o,t)$,and then the radiation field on the surface of receiver plane can be recorded as $E_r(x_r,t)$, and this propagation of the radiation field is denoted as transmitting path. In an actual engineering system, the light field in reference path $E_c(x_c,t)$ can be calculated based on the system configuration and detecting information, the effect of reference field is to eliminate the phase interference of system and propagation by coherence in essence. And then, the receiving radiation field and local oscillator($LO$) radiation field inputs into coherence detection and the output signal $i(t)$ can have first-order field correlation with the reference radiation field, which is organized by the calculating method in advance.
Assuming there is no signal modulated at the phase of the carrier, the radiation field $E_s(x_s,t)$ can be written as
\begin{equation}
E_{s}\left(x_{s}, t\right)=E_{s}\left(x_{s}\right) \exp \left(j 2 \pi f_{c} t\right)
\end{equation}
where $f_c$ is the frequency of carrier. After the propagation from the multiple transmitters to the single pixel coherent receiver, the radiation field on the receiver reflected from targets can be written as
\begin{figure*}[h!]
\centering
\includegraphics[width=2\columnwidth,height=0.55\linewidth]{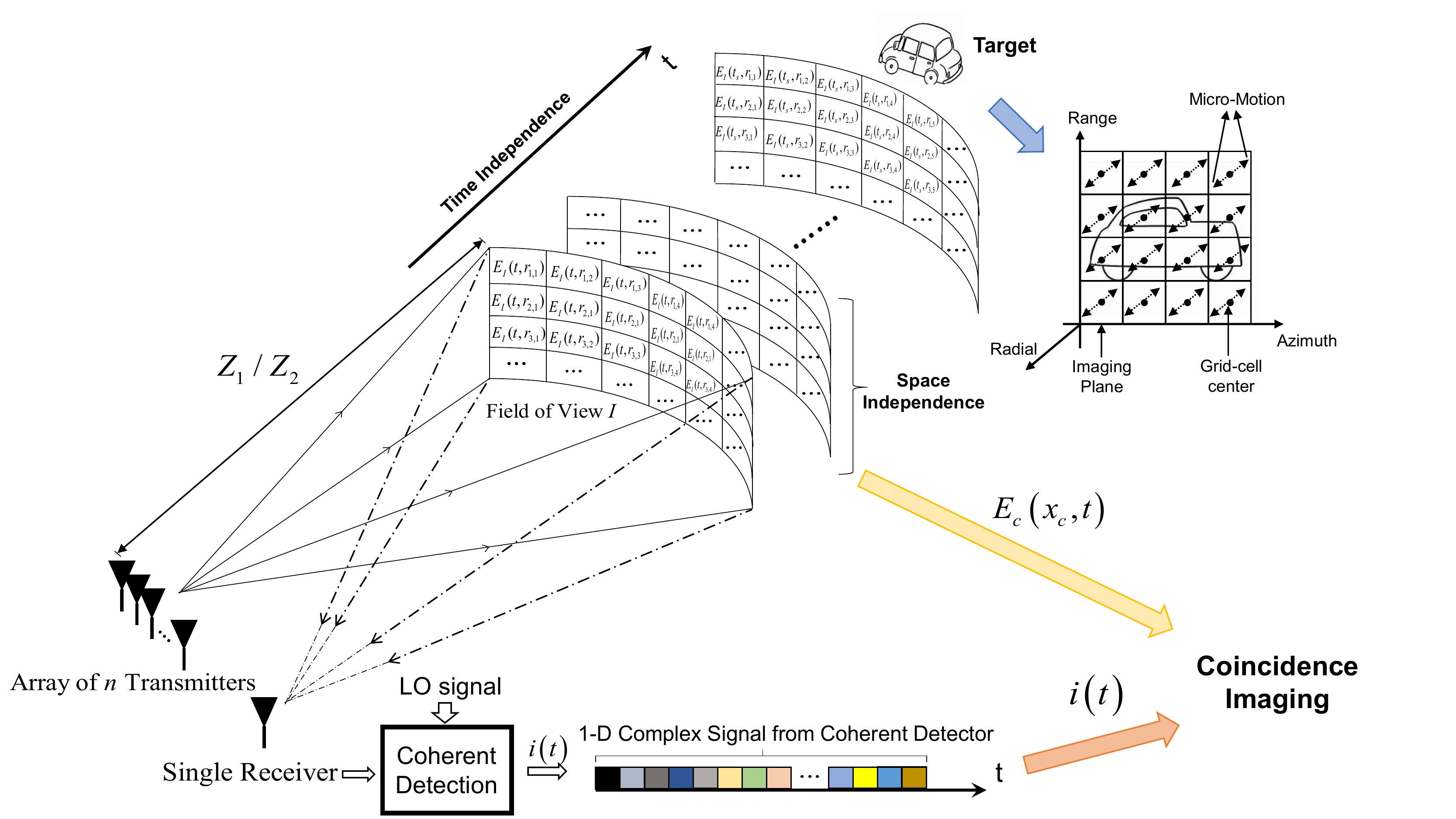}
\caption{The radar first-order field correlation system schematic diagram, considering wavelength order of the optical systems, the phase coherence of the optical field is affected by atmospheric turbulence and systematic errors during transmission. And the reference path radiation field can be calculated more accurately in the microwave systems, hence the radar systems are more suitable for the remote sensing scenarios }
\label{fig：1}
\end{figure*}
\begin{equation}
\begin{aligned}
E_{x_r}(t)=&\int_{x_{s}} d x_{s} E_{s}\left(x_{s}\right) h_{x_r}\left(x_{s}, t\right) \\
&\times\exp \left\{j 2 \pi f_{c}\left[t-\frac{Z_{1}(t)+Z_{2}(t)}{c}\right]\right\}
\end{aligned}
\end{equation}
In equation $(2)$, the distance items $Z_1$ and $Z_2$ are set as temporary variables $Z_1(t)$ and $Z_2(t)$ for universality, which are produced by the translation motions and micro vibrations, and $h_{x_r}(x_{s}, t)$ is the propagating function which can be written as\\

\begin{equation}
\begin{aligned}
h_{x_r}\left(x_{s}, t\right) & \propto \int_{x_{o}} d x_{o} \tilde{\mathcal{T}}\left(x_{o}, t\right) \exp \left[\frac{j k_{\lambda}}{2 Z_{1}(t)}\left(x_{o}-x_{s}\right)^{2}\right] \\
& \times \exp \left[\frac{j k_{\lambda}}{2 Z_{2}(t)}\left(x_{r}-x_{o}\right)^{2}\right]
\end{aligned}
\end{equation}

For a micro-vibration target model with a translation motion in receiving signal, we construct it as a complex valued target consisted of a reflection coefficient and a complex function caused by two types of motions, its detailed models will be elucidated in the Section III. The coherent detector outputs one-dimensional electronic signal $i(t)$ after the coherence between $E_{x_r}(t)$ and the $E_{LO}(t)$, 
\begin{equation}
i(t)=\eta  E_{x_r}\left( t\right) \cdot E_{L O}^{*}(t)
\end{equation}
where $\eta$ is the receiving efficiency of the coherent single pixel detector. In a realistic engineering system, the background noise is a low-frequency noise, the frequency shift can ensure the received signal away from the influence of the low-frequency noise. Thus the $LO$ field output by the frequency shifter can be expressed as
\begin{equation}
E_{L O}(t)=A_{L O} \exp \left[j 2 \pi\left(f_{c}-f_{IF}\right) t\right]
\end{equation}
where $f_{IF}$ is the shifting frequency. Thus the electronic signal $i(t)$ which is the coherent result between the $E_{LO}(t)$ and $E_{x_r}(t)$ can be expressed as

\begin{equation}
\begin{aligned}
i(t)&=\eta  A_{L O} \int_{x_{s}} d x_{s} E_{s}\left(x_{s}, t\right) h_{x_r}\left(x_{s}, t\right) \\
&\quad \times \exp \left[j 2 \pi f_{I F} t\right] \exp \left[-j 2 \pi f_{c} \frac{Z_{1}(t)+Z_{2}(t)}{c}\right] \\
&\quad \propto \int_{x_{s}} d x_{s} E_{s}\left(x_{s}, t\right) h_{x_r}\left(x_{s}, t\right) \\
  &\quad\times\exp \left[-j 2 \pi f_{c} \frac{Z_{1}(t)+Z_{2}(t)}{c}\right]
\end{aligned}
\end{equation}

For the output signal from coherent detector $i(t)$, the signal will be first-order correlated with the target surface radiation field $E_c(x_c,t)$, which can be calculated based on the prior knowledge of this detecting signal spatial distribution\cite{JSCI}. $E_c(x_c,t)$ is known as the reference path field, which can be written as
\begin{equation}
E_{c}\left(x_{c}, t\right)=\int_{x_{s}} d x_{s} E_{s}\left(x_{s}\right) h_{c}\left(x_{s}, x_{c}\right) \exp \left[j k_{\lambda} Z_{1}(t)\right]
\end{equation}
where $h_c(x_s,x_c)$ represents the propagation function from the source plane to the target plane, it can be written as
\begin{equation}
h_{c}\left(x_{s}, x_{c}\right) \propto \exp \left[\frac{j k_{\lambda}}{2 Z_{1}(t)}\left(x_{s}-x_{c}\right)^{2}\right]
\end{equation}

By this stage, the first-order correlation function associated with the reference path radiation field in equation $(7)$ and the one-dimensional electronic signal in equation $(6)$ can be given as 
\begin{equation}
\begin{aligned}
&G^{(1)}_{x_r}\left(x_{c}, t\right)\\
&=\left\langle E_{c}^{*}\left(x_{c}, t\right) i(t)\right\rangle = \eta \left\langle E_{c}^{*}\left(x_{c}, t\right) E_{x_r}(t)\cdot E_{L O}^{*}(t)\right\rangle \\
&\propto\left\langle\begin{array}{l}
\int_{x_s^{\prime}} d x_s^{\prime} \int_{x_{s}} d x_{s} E_{s}\left(x_{s}\right) E_{s}\left(x_{s}^{\prime}\right) \\
\times h_{c}^{*}\left(x_{s}, x_{c}\right) h_{x_r}\left(x_{s}, t\right) \exp \left[-j k_{\lambda} Z_{2}(t)\right]
\end{array}\right\rangle
\end{aligned}
\end{equation}
where $G^{(1)}$ refers to the first-order correlation function of a fluctuating scalar wavefield, and this function analyzes coherence phenomena of two radiation field $E_c(x_c,t)$ and $E_{x_r}(t)$ in essence. At the present stage, there are different theories about the expression of the first-order field correlation function. For higher order correlations, some theories refer to $G^{(M,N)}$ as the (space-time) cross-correlation function of order $(M,N)$ of radiation fields, correlation functions for which $N=M$ are particularly useful. They are often referred to as correlation functions of order $2M$. This nomenclature is inconsistent because some authors\cite{HOC} refer to $G(M,M)$ as a correlation function of order $M$ as opposed to 2$M$. In this paper, we will use $G^{(1)}$ to represent the expression of the first-order correlation function. Substituting equations $(3)$ and $(8)$ into equation $(9)$
\begin{equation}
\begin{aligned}
&G^{(1)}_{x_r}\left(x_{c}, t\right)\\
&\propto\left\langle\begin{array}{l}
\exp \left[-j k_{\lambda} Z_{2}(t)\right] \int_{x_{s}} d x_{s}^{\prime} \int_{x_{s}} d x_{s} E_{s}\left(x_{s}\right) E_{s}^{\prime}\left(x_{s}\right) \\
\times \int_{x_{o}} d x_{o} \widetilde{\mathcal{T}}\left(x_{o}, t\right) \exp \left[\frac{j k_{\lambda}}{2 Z_{1}(t)}\left(x_{o}^{2}-x_{c}^{2}\right)\right] \\
\times \exp \left[\frac{j k_{\lambda}}{2 Z_{1}(t)}\left(x_{c}-x_{o}\right) 2 x_{s}\right] \\
\times \exp \left[\frac{j k_{\lambda}}{2 Z_{2}(t)}\left(x_{r}-x_{o}\right)^{2}\right]
\end{array}\right\rangle
\end{aligned}
\end{equation}

As items in equation $(10)$ are temporal varying, the result of the first-order correlation would obscure the complex values after ensemble average, the information of a complex target cannot reconstruct precisely. We assume that the target is a perfectly stationary model in order to discuss the time-varying target model. In this scenario, the target model can be rewritten as a static complex target $\widetilde{\mathcal{T}}(x_{o})$, and meanwhile it is assumed that the relative distance between the target and the detection system remains constant, which means $Z_1(t)=Z_1,Z_2(t)=Z_2$. Then the first-order correlation function can be rewritten under these conditions as
\begin{equation}
\begin{aligned}
&G^{(1)}_{x_r}\left(x_{c}\right) \\
&\propto\left\langle\begin{array}{l}
\exp \left(-j k_{\lambda} Z_{2}\right) \int_{x_{s}} d x_{s}^{\prime} \int_{x_{s}} d x_{s} E_{s}\left(x_{s}\right) E_{s}^{\prime}\left(x_{s}\right) \\
\times \int_{x_{o}} d x_{o} \widetilde{\mathcal{T}}\left(x_{o}\right) \exp \left[\frac{j k_{\lambda}}{2 Z_{1}}\left(x_{o}^{2}-x_{c}^{2}\right)\right] \\
\times \exp \left[\frac{j k_{\lambda}}{2 Z_{1}}\left(x_{c}-x_{o}\right) 2 x_{s}\right] \\
\times \exp \left[\frac{j k_{\lambda}}{2 Z_{2}}\left(x_{r}-x_{o}\right)^{2}\right]
\end{array}\right\rangle
\end{aligned}
\end{equation}Assuming that the radiation field source is spatially completely incoherent and uniform in free space, then the first-order correlation function $G^{(1)}(x_s,x_{s}^{\prime})$ on the source surface can be expressed by the Dirac function\cite{FO}:
\begin{equation}
G^{(1)}\left(x_{s}, x_{s}^{\prime}\right)=I_{0} \delta\left(x_{s}-x_{s}^{\prime}\right)
\end{equation}
where $I_0$ stands for the intensity of the radiation field. Meanwhile, the first order of the light source's integration can be expressed as 
\begin{equation}
\int_{x_{s}} d x_{s} \exp \left[\frac{j k_{\lambda}}{2 Z_{1}}\left(x_{c}-x_{o}\right) 2 x_{s}\right]=\operatorname{sinc}\left[\frac{D}{\lambda Z_{1}}\left(x_{c}-x_{o}\right)\right]
\end{equation}
where $D$ stands for the equivalent aperture of the multiple transmitters. Under the hypothesis of ideal resolution, surface details of target can be perfectly mapped to the reference plane. Based on the condition of ideal resolution, the target plane coordinate system $x_o$ could be replaced with the reference path field coordinate system $x_c$. Substituting equations $(12)$ and $(13)$ into equation $(11)$,\\
\begin{equation}
G^{(1)}_{x_r}\left(x_{c}\right) \propto \exp \left(-j k_{\lambda} Z_{2}\right) \widetilde{\mathcal{T}}\left(x_{c}\right) \exp \left[\frac{j k_{\lambda}}{2 Z_{2}}\left(x_{r}-x_{c}\right)^{2}\right]
\end{equation}

It can be seen that except for the target complex valued item, there are some residual phase-items produced by the field propagation from equation $(14)$. The virtual parameters modulation can be introduced into reference path field to eliminate these extra phase-items, and the virtual modulated reference field can be rewritten as
\begin{equation}
E_{x_r,c}\left(x_{c}, t\right)=E_{c}\left(x_{c}, t\right) \widetilde{U}_{x_r}\left( x_{c}\right)
\end{equation}
where $\widetilde{U}_{x_r}\left( x_{c}\right)$ is a virtual compensation function which is computationally attached to the reference path. The purpose of the function $\widetilde{U}_{x_r}\left( x_{c}\right)$ is to filter the complex phase surface of the target by compensation and its form can change according to system configuration, thus the first-order field correlation function can be rewritten as $G^{(1)}_{x_r}\left(x_{c}, t\right)=\langle E_{x_r}^{*}( x_c,t) i(t)\rangle$.\\
For a static target detection, $\widetilde{U}_{x_r}\left( x_{c}\right)$ can be expressed as
\begin{equation}
\widetilde{U}_{x_r}\left( x_{c}\right)=\exp \left(j k_{\lambda} Z_{2}\right) \exp \left[-\frac{j k_{\lambda}}{2 Z_{2}}\left(x_{r}-x_{c}\right)^{2}\right]
\end{equation}
Thus substituting the new reference field function into first-order correlation function, the static complex target can be reconstructed perfectly as following
\begin{equation}
G^{(1)}\left(x_{c}\right) \propto \widetilde{\mathcal{T}}\left(x_{c}\right)
\end{equation}
According to equation $(17)$, the static target complex plane information can be reconstructed by calculating the first-order correlation function between the radiation field in reference path and the one-dimensional electronic signal. Consequently, a correlation imaging system for complex valued target phase restoration is presented in this section. Based on the theory of correlation imaging, the feasibility of this imaging scheme is demonstrated theoretically. It also provides a theoretical basis for the reconstruction of complex valued object with dynamic change due to vibration. \\
Since light belongs to electromagnetic (EM) waves in essence and there is no corpuscular property involved in the reconstruction procedures, this method can be further adopted and introduced into microwave imaging applications. A schematic diagram of a microwave coincidence radar is shown in Fig. 3, and since the phase item of the microwave can be controlled and recorded precisely, it is easier for the first-order field correlation to be realized in a microwave radar system than an optical system.

\section{Echo model based on the target micro-vibration model}

\begin{figure}[htbp]
\centering
\includegraphics[width=1.1\columnwidth,height=0.65\linewidth]{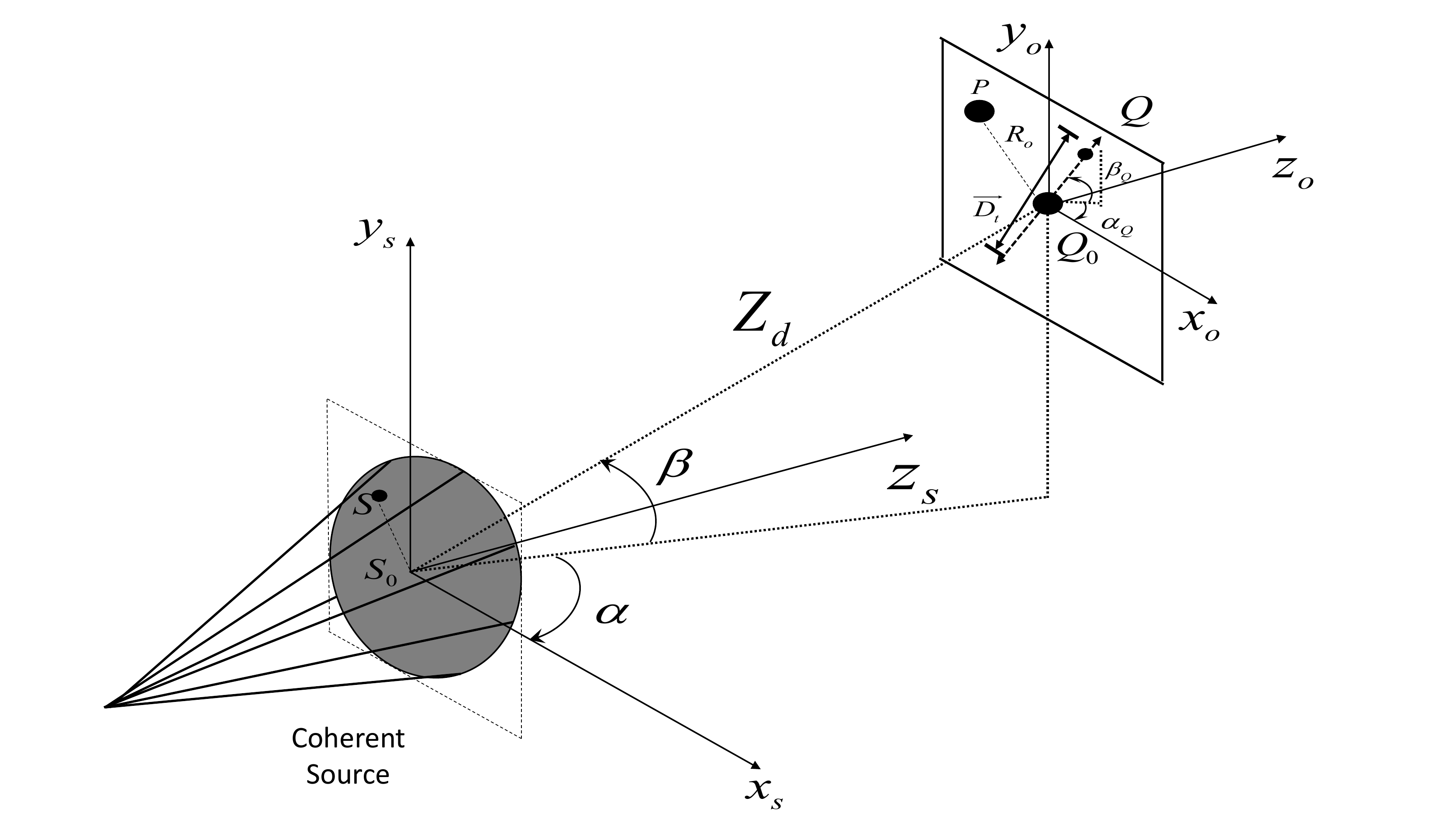}
\caption{Geometric diagram of target vibration and radar system, the radar source plane and target plane is set as a pair of parallel planes, and the receiver is also located on the plane of the source }
\label{fig：1}
\end{figure}
The target is roughly modeled as a complex valued model $\widetilde{\mathcal{T}}(x_{c},t)$ in the prior section, it is necessary to build the analytic expression of the target model for the further micro-vibration characteristics research. The target model based on the echo model is elucidated in this section. Another condition needs to be illustrated is that the discussion about micro-vibration model is completed in three dimensional coordinate in this section. Differing from the other sections of expressing a two-dimensional plane in terms of one-dimensional variable, this section uses three variables $x,y,z$ to represent three-dimensional changes.\\
As illustrated in Fig. 4, the radar source plane is in plane  $x_s S_0 y_s$ of a stationary spatial coordinate system $(x_s,y_s,z_s)$ while the target is in plane $x_o Q_0 y_o$ of the reference coordinate system $(x_o,y_o,z_o)$ with a distance $Z_d$ from the center of the source. Assuming that the center of the reference coordinate system $Q_0$ is also the center of the target plane. And there is a scattering point $Q$ in this plane, the point $Q$ vibrates periodically in a certain direction and the vibration's center location is $Q_0$. $\alpha, \beta$ are the azimuth and elevation angle of radar respectively, in the same way, $\alpha_Q, \beta_Q$ are the azimuth and elevation angle of the vibration direction of the vibration scattering center in the reference coordinate system. Assuming that the distance between $Q_0$ and radar is $\left|\vec{Z}_{d}\right|=Z_{d}=Z_{1}=Z_{2}$, thus the position of the vibration center in coordinate system $(x_s,y_s,z_s)$ can be represented as \\
\begin{equation}
\left(Z_d \cos \beta \cos \alpha, Z_d \sin \beta, Z_d \cos \beta \sin \alpha\right)
\end{equation}
Furthermore, $\eta(t)$ represents the distance periodical change of point Q, at time $t$, the initial distance between $Q$ and $Q_0$ is $\eta_t$, then the instantaneous coordinate of $Q$ in the reference coordinate system is
\begin{equation}
\left(\eta_{t} \cos \beta_{Q} \cos \alpha_{Q}, \eta_{t} \sin \beta_{Q}, \eta_{t} \cos \beta_{Q} \sin \alpha_{Q}\right)
\end{equation}
The distance vector from the radar to the scattering point $Q$ is $\overrightarrow{Z_{t}}=\overrightarrow{Z_d}+\overrightarrow{\eta_{t}}$, and in a stationary radar coordinate system, the instantaneous value of the distance can be expressed as:
\begin{equation}
Z_{t}=\left|\overrightarrow{Z_{t}}\right|=\left[\begin{array}{l}
\left(Z_d \cos \beta \cos \alpha+\eta_{t} \cos \beta_{Q} \cos \alpha_{Q}\right)^{2} \\
+\left(Z_d \sin \beta+\eta_{t} \sin \beta_{Q}\right)^{2} \\
+\left(Z_d \cos \beta \sin \alpha+\eta_{t} \cos \beta_{Q} \sin \alpha_{Q}\right)^{2}
\end{array}\right]^{\frac{1}{2}}
\end{equation}
On account of $z_{0} \gg \eta_{t}$, in order to keep the micro-vibration item, the equation $(20)$ can be simplified as
\begin{equation}
\begin{aligned}
Z_{t}&=\left[\begin{array}{l}
Z_d^{2}+\eta_{t}^{2}+2 Z_d \eta_{t}\left(\cos \beta \cos \alpha \cos \beta_{Q} \cos \alpha_{Q}\right. \\
\left.+\sin \beta \sin \beta_{Q}+\cos \beta \sin \alpha \cos \beta_{Q} \sin \alpha_{Q}\right)
\end{array}\right]^{\frac{1}{2}} \\
&\approx Z_d+\eta_{t}\left[\cos \beta \cos \beta_{Q} \cos \left(\alpha-\alpha_{Q}\right)+\sin \beta \sin \beta_{Q}\right]
\end{aligned}
\end{equation}The vector representation of target translation $v$ is discussed below: assuming that $\theta$ and $\gamma$ are the azimuth and elevation angle of the translation direction in the stationary coordinate system respectively, and the translation vector can be written as
\begin{equation}
\vec{v(t)}=(vt \cos \gamma \cos \theta, v t \sin \gamma, v t \cos \gamma \sin \theta)
\end{equation}
At this point, the instantaneous distance between scattering point $Q$ and radar source center $S_0$ can be rewritten as\\
\begin{equation}
Z(t)=Z_d+v(t) t+\eta_{t}\left[\cos \beta \cos \beta_{Q} \cos \left(\alpha-\alpha_{Q}\right)+\sin \beta \sin \beta_{Q}\right]
\end{equation}
The motion model needs to be solved in the whole field of view, based on the single point $Q$ motion, the scattering point $Q$ can be expanded to the detecting area, including the two-dimensional plane and the three-dimensional depth.\\
Because the radiation field in free space propagates as a spherical wave, every point on a two-dimensional plane with a center point at $Z_d$ from the center of the source plane may have a phase difference with the center point. Phase difference caused by the transverse distance can be compensated by the propagating function $H(x_s,y_s,x_o,y_o)$ between the source plane and the  target plane, its expression is
\begin{equation}
H\left(x_{s}, y_{s}, x_{o}, y_{o}\right)=\exp \left\{\frac{j k_{\lambda}}{2 Z_d}\left[\left(x_{s}-x_{o}\right)^{2}+\left(y_{s}-y_{o}\right)^{2}\right]\right\}
\end{equation}
The target can be seen as a rigid body for translation motion when the axial rotations do not exist, all of the scattering points of the whole target remain the same translation state. The relative function that characterizes the time-varying phase difference between scattered points is set as $w(x_o, y_o, z_o,t)$ for micro-vibration motions. Meanwhile, the angle function is set as $A(\alpha,\beta,\alpha_Q,\beta_Q)=\cos \beta \cos \beta_{Q} \cos \left(\alpha-\alpha_{Q}\right)+\sin \beta \sin \beta_{Q}$, and the distance function from each points on the target plane to the center of source plane can be rewritten as\\
\begin{equation}
Z(x_o, y_o, z_o, t)=Z_d+v(t) t+w\left(x_{o}, y_{o}, z_{o}, t\right)A(\alpha,\beta,\alpha_Q,\beta_Q)
\end{equation}

The single pixel coherent receiver is located on the same plane of the source. In the microwave sensing system, the distance between receiver and source center can be ignored, thus the phase difference caused by the distance between the receiver and the transmitters can be ignored, we can assume that the single pixel receiver is also located at the center of the source plane. Differing from conventional radar sensing systems, the radiation field generated by the transmitters has spatial modulated patterns, thus the radiation field on the target plane should have spatial coding characteristics, it is assumed that the patterns illuminated that field of view can be represented by the function $P(x_o,y_o,z_o,t)$ which can be written as\\
\begin{equation}
\begin{aligned}
P\left(x_{o}, y_{o}, z_{o}, t\right)&=\int d x_{s} E_{s}\left(x_{s}, t\right) \exp \left(j k_{\lambda} Z_d\right) \\
&\times H\left(x_{s}, y_{s}, x_{o}, y_{o}\right)
\end{aligned}
\end{equation}

Set $\phi(x_o, y_o, z_o, t)=4 \pi Z(x_o, y_o, z_o, t) / \lambda_{c}$ where $\lambda_c$ is the carrier wavelength, and the function $\phi(x_o, y_o, z_o, t)$ represents all distance-dependent phase changes. In the field of view $I$,  the output electronic signal of the receiving radiation field detected by a single receiver after coherence with the $LO$ field can be written as  
\begin{equation}
\begin{aligned}
&i(t)=\int_{I} d x_o d y_o d z_o \cdot P(x_o, y_o,z_o, t) \cdot \mathcal{T}(x_o, y_o, z_o)\\
&\times \exp \left\{j\left[2 \pi f_{I F} t+\phi(x_o, y_o, z_o, t)\right]\right\}\\
&\times H\left(x_{o}, y_{o}, x_{r}, y_{r}\right)+\omega(t)\\
&=\int_{I} d x_o d y_o d z_o \cdot \mathcal{T}(x_o, y_o, z_o)\\
&\times \exp \left\{\frac{j 4 \pi}{\lambda_{c}}\left[\begin{array}{l}
\frac {f_{I F} t} {2}+Z_d+v(t) t \\
+W(x_o, y_o, z_o, t) A\left(\alpha, \beta, \alpha_{Q}, \beta_{Q}\right)
\end{array}\right]\right\}\\
&\times H\left(x_{o}, y_{o}, x_{r}, y_{r}\right)+n(t)
\end{aligned}
\end{equation}where, $f_c$ is the carrier frequency , $c$ is the light speed, $n(t)$ is the background noise, $\mathcal{T}(x_o, y_o, z_o)$ is the scattering coefficient of the whole target plane, and $H(x_o,y_o,x_r,y_r)$ stands for the propagation function from the target plane to the receiving plane. According to the echo model, this paper will keep on our research based on these reduced conditions introduced below:
\begin{enumerate}
\item When both of the azimuth and elevation of a radar system are $0$, then $\alpha=0, \beta=0$.
\item When the azimuth and elevation of the scattering point $Q$ vibration direction are $0$ in the reference coordinate system, then $\alpha_Q=0,\beta_Q=0$.
\item This paper focuses on the micro-vibration modes spatial distribution and target micro-vibration characteristics recognition, thus it is assumed that the target translation is $0$, then $v(t)=0$.
\item To keep things simple, this study solely discusses two-dimensional targets and ignores target depth, hence $Z_d=0$.
\item  Ideally, background noise is not considered, thus $n(t)=0$.
\end{enumerate}
Based on these simplifications, the equation $(27)$ can be rewritten as\\
\begin{equation}
\begin{aligned}
i(t) &=\int_{I} d x_{o} d y_{o} \cdot P\left(x_{o}, y_{o}, t\right) \mathcal{T}\left(x_{o}, y_{o}\right) \\
& \times \exp \left\{\frac{j 4 \pi}{\lambda_{c}}\left[\frac{f_{I F} t}{2}+Z_d+w\left(x_{o}, y_{o}, t\right)\right]\right\} \\
& \times H\left(x_{o}, y_{o}, x_{r}, y_{r}\right)
\end{aligned}
\end{equation}
Consequently, according to equation $(28)$, except for the phase changes due to propagation, the target complex model is determined, assuming there are $k$ scattering points in the field of view, then the target complex model can be written as
\begin{equation}
\widetilde{\mathcal{T}}\left(x_{o, k}, t\right)=\mathcal{T}\left(x_{o}\right) \delta\left(x_{o}-x_{o, k}\right) \exp \left[j 2 k_{\lambda} w\left(x_{o, k}, t\right)\right]
\end{equation}
where, the Delta function stands for the location of the scattering points, and the detailed formation of vibration function $w(x_{o,k},t)$ are different in discrete targets and continuous targets, that will be discussed in the following section.

\section{Discussions about two types of target micro-vibration models }
The detailed models of different objects will be proposed in this section. The objective periodical vibration remains stable during the whole correlation detection in the realistic scenarios.
\begin{figure}[h!]
\centering
\includegraphics[width=1\columnwidth,height=0.5\linewidth]{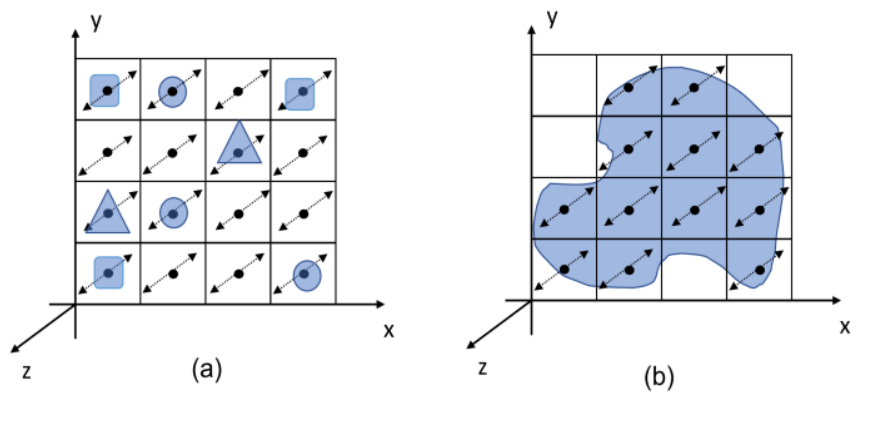}
\caption{Two types of targets schematic diagram, for discrete targets, the size is less than one field speckle resolution, and on the other side there will be multiple speckles illuminating the continuous target in the field of view. In essence, the continuous target is also composed of multiple scattering points, but since these scattering points are located on the same target surface, these scattering points have obvious spatial constraints, and their micro-vibration model is established under a transverse model framework }
\label{fig：1}
\end{figure}
Targets can be classified into two categories based on their size and the minimal resolution unit in the field of view: discrete targets and continuous targets. For simplicity, we assume that the target does not lie on the boundary between the two resolution units, which might lead to ambiguity in the final picture.  As shown in Fig. 5, when the single target size is smaller than a speckle in the field of view, this target can be seen as a scattering point, in this case, the micro-vibration of a single target has no spatial distribution. When the target is much greater than a single speckle scale, the micro-vibration characteristics of each scattering point on the target must conform to the constraint of function $w(x_o,t)$ . 

\subsection {Two-dimensional discrete micro-vibration model}
The objects to be detected are a group of scattered points that have no spatial relevance in essence in discrete targets detecting scenarios. Assuming that there are $k$ objects in the field of view, then 
\begin{equation}
w\left(x_{o,k}, t\right)=\sum_{k=1}^{K} w(x_o) \delta(x_o-x_{o,k}) Z_{k} \eta_{k}(t)
\end{equation}
where $Z_k$ stands for the amplitude of the $k$th vibrating objects. For the stable point periodical vibration, the micro-vibration function $\eta_{k}(t)$ can be expanded by the Fourier series. As a result, equation $(30)$ can be rewritten as
\begin{equation}
\begin{aligned}
w\left(x_{o,k}, t\right) &=\sum_{k=1}^{K} \sum_{n=0}^{\infty} w(x_o) \delta(x_o-x_{o,k})\\
&\times Z_{k, n} \cos \left(2 \pi  f_{k, n} t+\phi_{k, n}\right)
\end{aligned}
\end{equation}
Each of Fourier components has two constant items: amplitude $Z_n$ and initial phase $\varphi_{n}$. $f_n$ is schema of each vibration which can be detected in the  frequency spectrum after the Fourier transform. Due to the fact that the discrete target can be equivalent to several scattering points, the echo consists of multiple scattering information superimposed, when we have number $k$ discrete targets, then the $k$th target complex model can be rewritten as
\begin{equation}
\begin{aligned}
&\widetilde{\mathcal{T}}\left(x_{o, k}, t\right)\\
&=\sum_{k=1}^{K} \sum_{n=0}^{\infty} \mathcal{T}\left(x_{o}\right) \delta\left(x_{o}-x_{o, k}\right)  \\
&\times\exp \left[j 2 k_{\lambda}  w(x_o) \delta(x_o-x_{o,k}) Z_{k, n} \cos \left(2 \pi  f_{k,n} t+\phi_{k,n}\right)\right]
\end{aligned}
\end{equation}
%In a realistic scenario, it always has several discrete targets sharing the same vibration mode, by detecting a certain vibration mode, those targets which having the same mode can be reconstructed in the final image.

\subsection {Two-dimensional continuous sheet micro-vibration model}
\begin{figure}[h!]
\centering
\includegraphics[width=1\columnwidth,height=0.65\linewidth]{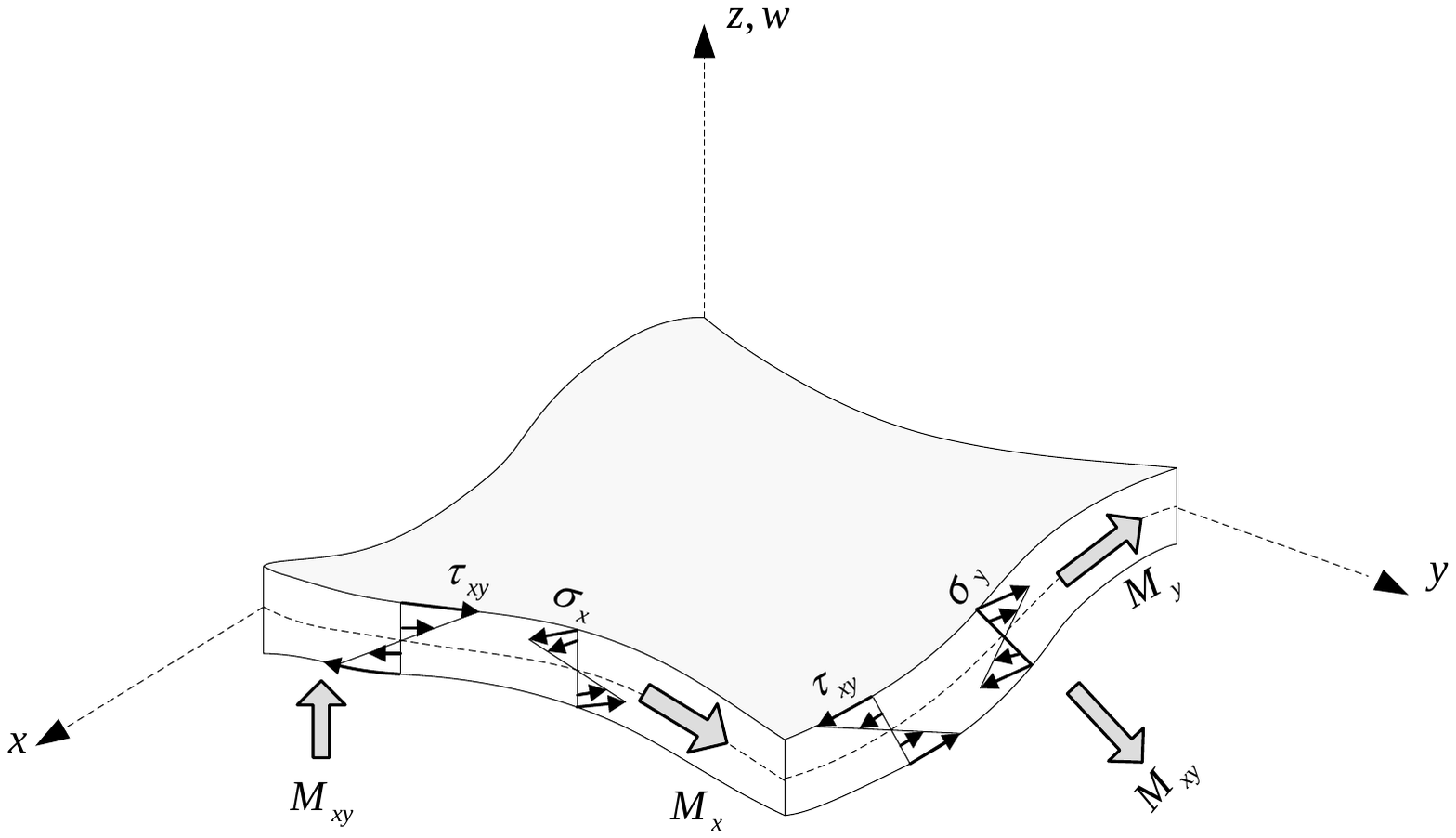}
\caption{The schematic diagram of micro vibrations and boundaries of continuous sheet model in 2-D plane are shown in this figure, where $\tau$ represents transverse shear stress of bending plate,  $\sigma$ represents internal stresses , and $M$ is for a moment in a 2-D plane.}
\label{fig：1}
\end{figure}
Unlike discrete objects, the strain characteristics of non-grid bodies should be taken into account. Each scattering point is determined not only by longitudinal vibration but also by transverse propagation, with longitudinal expansion accompanied by lateral contraction. There are two types of transverse propagation systems for the objects: stationary wave systems and traveling wave systems. Whereas the objects to be measured in most of our detecting scenarios are stationary wave systems, the traveling wave systems model will not be discussed in this paper.\\

The principal modes, also known as the transverse propagation distribution on the surface of the target, can be simplified into a one-dimensional superposition of $M$ space fundamental modes based on the separation of space and time. The theoretical support of two-dimensional continuous sheet model can be found in appendix. And this stationary wave model allows the continuous object model to be expressed as\\
\begin{equation}
\begin{aligned}
&\widetilde{\mathcal{T}}\left(x_{o}, t\right)=\mathcal{T}\left(x_{o}\right) \exp \left[j  2 k_{\lambda} w\left(x_{o}, t\right)\right] \\
&=\sum_{m=0} ^{\infty}\mathcal{T}_m\left(x_{o}\right)  \exp \left[j 2 k_{\lambda} W_{m}\left(x_{o}\right) \eta_{m}(t)\right]
\end{aligned}
\end{equation}

Every vibration has its corresponding space mode, even though each space mode normally corresponds to a single frequency vibration for realistic targets. Without loss of generality, the periodical vibration function $\eta_{m}(t)$ can also be expanded by the $n$ order Fourier series, thus formula $(33)$ can be further rewritten as,
\begin{equation}
\begin{aligned}
\widetilde{\mathcal{T}}\left(x_{o}, t\right) &=\sum_{m=0}^{M}  \mathcal{T}_{m}\left(x_{o}\right) \\
&\times \exp \left[j 2 k_{\lambda} W_{m}\left(x_{o}\right) \sum_{n=0}^{\infty} \cos \left(2 \pi f_{m n} t+\phi_{m n}\right)\right]
\end{aligned}
\end{equation}
where $\mathcal{T}_{m}\left(x_{o}\right) $ is the reflectivity image of the $M$th principal mode. The constraint models of each reflection point in two dimensional continuous target plane have been established explicitly from function $(34)$. The whole target is composed of $M$ principal modes and each mode corresponds a vibration time varying function. A time-varying function can be fitted by the Fourier series without loss of generality. According to the target model, all the interesting information is contained in the phase item, consequently, the phase reconstruction by field correlation function provides theoretical support for further extraction of these mode information.

\section {Micro-Vibration Modes Detection Algorithm}
Based on the target models introduced previously, this section proposes the methods to extract vibrating target time-spatial characteristic variables. The phase information including the vibration modes is generally lost in traditional ghost imaging algorithms, and the phase information is not constant during correlation detection for vibrating objects. Thus the core problem in the micro-vibration modes detection is how to smooth and eliminate the phase item time varying feature during the detection.
\subsection {Discrete targets reconstructing derivation}
For discrete targets shown in equation $(32)$, substituting the discrete target model into the first-order field correlation function shown in equation $(10)$ 
\begin{equation}
\begin{aligned}
&G^{(1)}_{x_r}\left(x_{c}, t\right) \\
&\propto\left\langle\begin{array}{l}
\exp \left(-j k_{\lambda} Z_{2}\right) \int_{x_{s}} d x_{s}^{\prime} \int_{x_{s}} d x_{s} \int_{x_{o}} d x_{o} E_{s}\left(x_{s}\right) E_{s}^{\prime}\left(x_{s}\right) \\
\times  \sum_ {k = 1} ^{ K }  \sum_{n = 0}^{\infty} \int_{x_{o}} d x_{o} \mathcal{T}\left(x_{o} \right) \delta(x_o - x_{o,k})  \\
\times  \exp \left[j 2 k_{\lambda}  w(x_o) \delta(x_o-x_{o,k}) %\\
\times Z_{k, n} \cos \left(2 \pi f_{k, n} t+\phi_{k, n}\right)\right]\\
\times  \widetilde{U}_{x_r}\left( x_{c}\right) \exp \left[\frac{j k_{\lambda}}{2 Z_{1}(t)}\left(x_{o}^{2}-x_{c}^{2}\right)\right]  \\
\times \exp \left[\frac{j k_{\lambda}}{2 Z_{1}(t)}\left(x_{c}-x_{o}\right) 2 x_{s}\right] \exp \left[\frac{j k_{\lambda}}{2 Z_{2}(t)}\left(x_{r}-x_{o}\right)^{2}\right]
\end{array}\right\rangle
\end{aligned}
\end{equation}

From the preceding analysis, it can be seen that the characteristic curves extraction algorithm can hardly recover different objects motion mode, but the frequencies of vibration modes could be acquired readily. Considering that the prior vibration model for discrete objects have established, the virtual time coding which is modulated in the speckle field of the reference path can be modified to eliminated the target time varying micro-vibration. For a certain micro-vibration mode $\epsilon$, its vibration frequency $f_{\epsilon,n}$, amplitude $Z_{\epsilon,n}$ and initial phase $ \phi_{\epsilon,n} $ are the parameters can be found in the time-frequency spectrum, and their corresponding spatial distribution should be $\mathcal{T}\left(x_{o, \epsilon}\right)$. To extract the target micro-vibration spatial distribution image, this method needs to eliminate the phase vibration characteristics of the target and make the first-order field correlation, so that the target reflectance image will not be blurred with the ensemble averaging process. Based on the virtual modulation code function $\widetilde{U}_{x_r}(x_c)$ which is introduced in equation $(15)$, we can rewrite the function $(16)$ on the basis of the discrete target model as,
\begin{equation}
\begin{aligned}
\widetilde{U}_{x_r}\left(x_{c}\right) &=\exp \left(j k_{\lambda} Z_{2}\right) \exp \left[-\frac{j k_{\lambda}}{2 Z_{2}}\left(x_{r}-x_{o}\right)^{2}\right] \\
& \times \exp \left[-j 2 k_{\lambda} Z_{\epsilon, n} \cos \left(2 \pi f_{\epsilon, n} t+\phi_{\epsilon, n}\right)\right]
\end{aligned}
\end{equation}

Then substituting equation $(36)$ into function $(35)$, and arranging the first-order field correlation function in order,
\begin{equation}
\begin{aligned}
&G^{(1)}\left(x_{c}, t\right) \propto \mathcal{T}\left(x_{c, \epsilon}\right) \\
&+\left\langle\begin{array}{l}
\sum_{k=1}^{K} \sum_{n=0}^{\infty} \mathcal{T}\left(x_{o, k}\right) \\
\times \exp \left[j 2 k_{\lambda} Z_{k, n} \cos \left(2 \pi f_{k, n} t+\phi_{k, n}\right)\right] \\
\times \exp \left[-j 2 k_{\lambda} Z_{\epsilon, n} \cos \left(2 \pi f_{\epsilon, n} t+\phi_{\epsilon, n}\right)\right]
\end{array}\right\rangle, k \neq \epsilon
\end{aligned}
\end{equation}
It can be seen from equation $(37)$ that, for the spatial distribution of target vibration mode $f_\epsilon$, the relevancy are accumulated within the multiple sampling matches. Meanwhile, despite the fact that other modes have projection residues on the target mode due to their non-orthogonality, these residual items will be eliminated under the condition of ensemble average, therefore the field correlation function can be reformulated as follows:\\
\begin{equation}
G^{(1)}\left(x_{c}\right) \propto \mathcal{T}\left(x_{c, \epsilon}\right)
\end{equation}

Since then, the theoretical discrete target model of the spatial distribution reconstruction algorithm has established from the result of equation $(38)$. It can be seen that when different discrete targets share the same model, multiple scattering points can be seen in the final image, this can help the imaging system to filtrate the collective targets based on their micro-vibration modes in the field of view.
\subsection {Discrete target reconstructing simulation results}
This method can be used in microwave radar systems whose emitting radiation field can be recorded and controlled by the multi-transmitters as discussed before\cite{MGI3,GFKD}. For simulation, the systems still set as an optical system on the condition that the light complex field is the prior value. The parameters of light systems are set as Table. \uppercase\expandafter{\romannumeral1}.\\
\begin{table}[h!]
\centering
\caption{PARAMETER SETTING FOR OPTICAL SYSTEM}
 %\begin{tabular}{|c{3cm} | c{8cm} |} 
 \begin{tabular}{|c | c|} 
 \hline
 Lambda & $\lambda = 1550e^{-9}$m\\ 
 \hline
 Period & T=1s \\ 
 \hline
 Sampling Rate & $f_s = 20s^{-1}$\\
\hline
Coding Method & Hadamard\\
\hline
Pattern Size & $32 \times 32$\\
 \hline
 \end{tabular}
 \label{table:1}
\end{table}

Different coding methods have different imaging resolution, we adopt Hadamard as our transmitting radiation field pattern as shown in Table. \uppercase\expandafter{\romannumeral1}, the imaging resolution of this encoding form is equal to the size of each speckle. There are some studies on super-resolution coding that can use varying speckle sizes to break through the resolution limits of traditional coding methods\cite{RES1,RES2}.\\
For the generality, equation $(32)$ introduce that every vibration mode are expanded by $n$-order Fourier series which can fit periodic vibrations of all objects. However, the majority of practical targets only vibrate at a single frequency, thus we adjusted each vibration mode in the simulation to correspond to a single frequency.\\
\begin{table}[h!]
\centering
\caption{PARAMETER SETTING FOR DIFFERENT VIBRATION MODES}
 \begin{tabular}{|c|c|c|c|} 
 \hline
 Parameters & Mode 1 & Mode 2 & Mode 3 \\ 
 \hline
 Amplitude(m) & 100e-9 & 500e-9 & 1200e-9 \\ 
 \hline
 Initial Phase(rad) & 0 & $\pi/4$ & $5\pi/6$ \\
 \hline
 Frequency(Hz) & 10 & 5 & 15 \\
 \hline
 \end{tabular}
 \label{table: 1}
\end{table}

Table.\uppercase\expandafter{\romannumeral2} shows parameters for the three vibration modes. Assuming that twelve objects randomly appear in the field of view, and each of them has the same reflection coefficient. As shown in Fig. 7, all of the targets only have a single reflection points in the field of view.\\
 \begin{figure}[h!]
\centering
\includegraphics[width=0.75\columnwidth,height=0.75\linewidth]{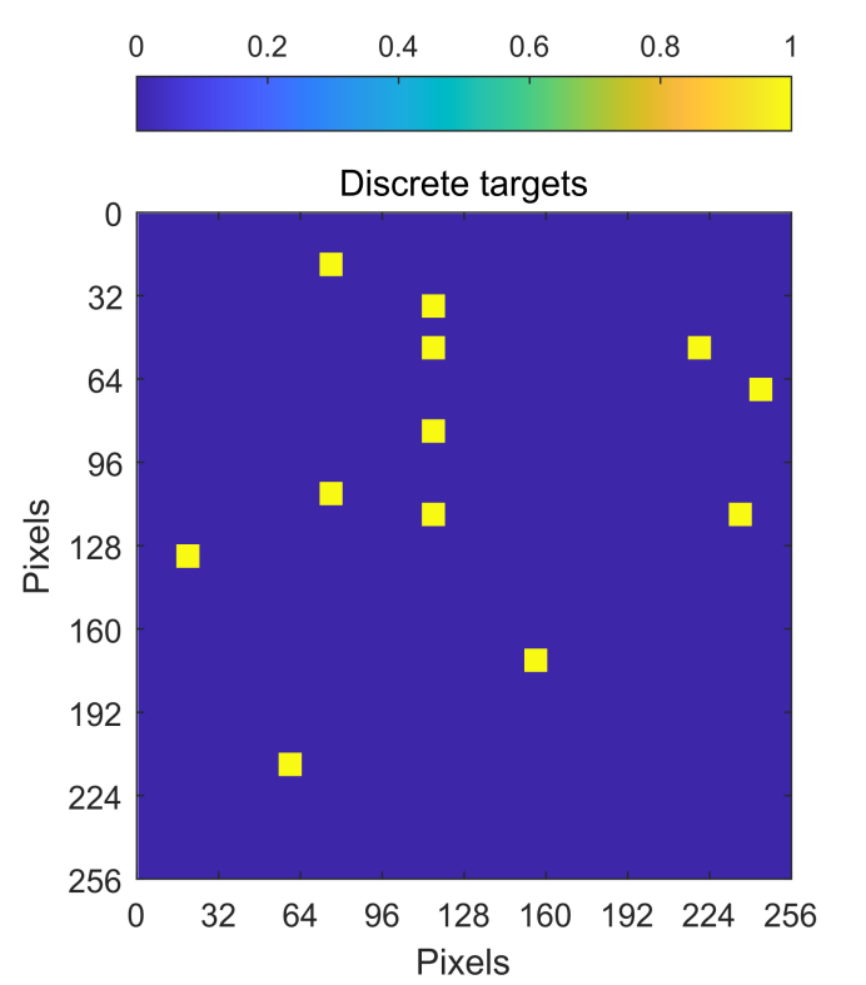}
\caption{The field of view has several discrete points targets and their scale is smaller than the resolution of speckle field transmitted to the target plane, the different vibration modes is remaining among them and the objects which owns the same vibration modes can be seen as the collective objects. In this scenario, these objects will be set to three modes randomly  }
\label{fig：1}
\end{figure}
And even though equation $(38)$ explains that the residual items will be eliminated under the condition of an ensemble average of infinite sampling, in realistic sensing systems, these residual items will be reserved due to the limited samples. Then a spatial relevance high-pass filter can be set on the final image, and these residual items can be perfectly filtered due to their non-correlation with virtual modulation function $\widetilde{U}_{x_r}\left( x_{c}\right)$.\\

 \begin{figure}[h!]
\centering
\includegraphics[width=1\columnwidth,height=0.8\linewidth]{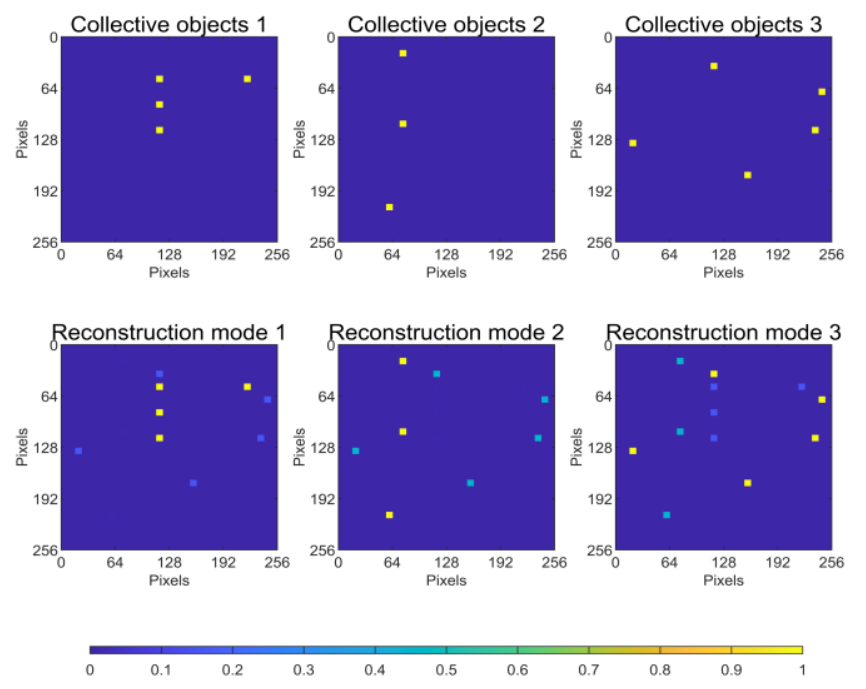}
\caption{The first row shows the original space distribution of collective objects respectively. And the second row shows the reconstruction result after time coherence which is filtered by the high correlation pass filter.  }
\label{fig：1}
\end{figure}
As shown in Fig. 8, it shows that several discrete targets sharing the same vibration mode, by detecting a certain vibration mode, those targets which having the same mode can be reconstructed in the final image. And different vibrations are filtered under the different virtual reference path modulating functions $\widetilde{U}_{x_r}\left(x_{c}\right)$. Each mode reconstruction result has redundancy in each target edge, which is caused by the speckle field patterns modulated in the reference path. Based on the static meshing method of speckle field, when the target place is in the boundary between the multiple subgrids of the speckle field, the target can be illuminated by several subgrids in one pulse, rather than by a single speckle spot. The boundary reconstruction effect of the target creates redundancy in that case. Except for the fixed error generated by the system, the reconstruction effect of each vibration mode is ideal in the absence of background noise.

\subsection {Continuous model reconstructing derivation}
\begin{figure}[h!]
\centering
\includegraphics[width=1.0\columnwidth,height=0.7\linewidth]{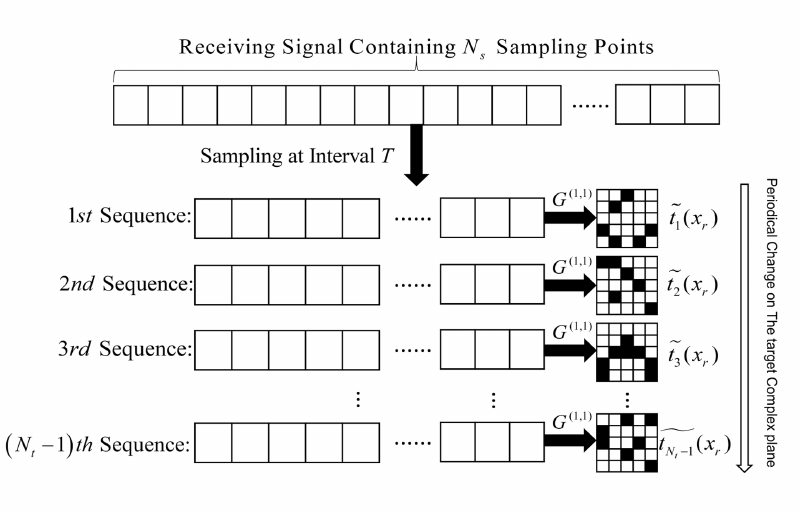}
\caption{Diagram about interval sampling method. For the sampling interval time $T$, this time interval is determined by all of the micro motion modes period time, and $T$ should be least common multiple of these modes. After the field correlation calculation, a sequence of images is acquired and this sequence can show the two-dimensional target plane's exact change in a period of time $T$.}
\label{fig：1}
\end{figure}

To begin, notice that every space distribution mode, also known as the principal mode, has its own one-dimensional time sequence vibration function in the two-dimensional continuous micro-vibration stationary wave model illustrated in equation $(34)$. Bringing the continuous target model up as follow,
\begin{equation}
\begin{aligned}
\widetilde{\mathcal{T}}\left(x_{o}, t\right) &=\sum_{m=0}^{M}  \mathcal{T}_{m}\left(x_{o}\right) \\
&\times \exp \left[j 2 k_{\lambda} W_{m}\left(x_{o}\right) \sum_{n=0}^{\infty} \cos \left(2 \pi f_{m n} t+\phi_{m n}\right)\right]
\end{aligned}
\end{equation}
The principal mode $W_m(x_o)$ can be interpreted as the spatial amplitude distribution in the two-dimensional target plane during the vibration procession of the target. Each mode has a corresponding reflection coefficient image, $M$ spatial modal reflecting coefficient images constitute the final target reflecting coefficient image $\mathcal{T}(x_o)$. The virtual time domain coding method introduced in the discrete targets vibration mode detection cannot be used in two-dimensional continuous target sensing scenarios. The continuous target can be seen as a collection of discrete scattering points constrained by a  transverse propagation function. There would be only several strong scattering points matched in the correlation with the virtual time coding reference path in continuous target detection, it is meaningless for the reconstruction of the principal modes. Since the principal mode function $W_m(x_o)$ of the signature objects has a distinct relevancy with material and shape characteristics, reconstructing the primary mode function $W_m(x_o)$ will greatly improve the ability to obtain high-dimensional information for a sensing system. Thus a new method is proposed in the continuous target detecting scenario, which can be called a time interval sampling method.\\

Because the first order field correlation function can reconstruct static complex target imaginary values, a sampling interval time $T$, which is also the period of target vibration, can be set in the echo sampling for a periodical vibration target. And a new series of time sequences will be acquired at the new sampling time interval. The complex target plane remains consistent in this time sequence. According to the continuous target model, the target plane has limited number principal modes. For an echo which has a long enough sampling time, we can always find a new time sequence sampled at the sampling interval $T$ from the original echo, and the target is equivalent to a static target in this new time sequence. For different values of sampling time interval $T$, there are two types of reconstruction methods to have different target images.\\
\begin{figure*}[h]
\centering
\includegraphics[width=2\columnwidth,height=0.35\linewidth]{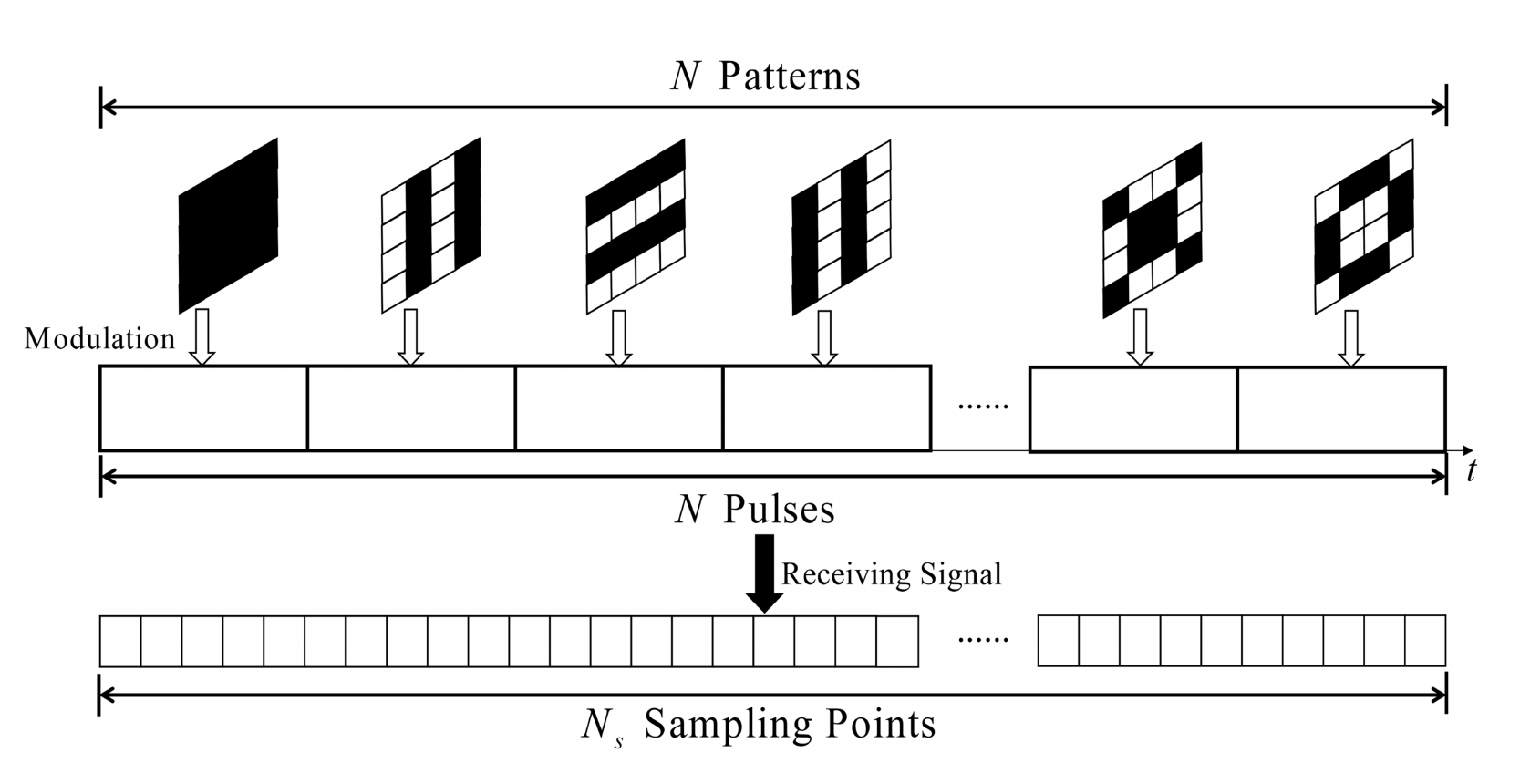}
\caption{The corresponding relation between $N$ speckle field patterns, $N$ transmitting pulses and $N_s$ sampling points. For the interval sampling method, the new time sequence interval sampled from the original echo need to find its matching speckle field pattern to process the field correlation. }
\label{fig：1}
\end{figure*}

$\boldsymbol{Type 1}$: Because function $\eta_m(t)$ represents the periodical vibration of the $M$th principal mode,  a sampling interval $N_t$ is set to eliminate space vibration characteristic and make the phase item of target plane remain consistent. To determinate the value of $N_t$, the frequencies of principal modes should be obtained first. As shown in the first branch of the flow diagram of Fig. 11, the time-frequency spectrum is acquired after the time frequency analysis. The time-frequency spectrum shows ${m}\times{n}$ time-frequency characteristic curves, which corresponds to its own vibration frequency $f_{m n}$, overlapping in the spectrum. The period time of each vibration frequency $f_{m n}$ is denoted as $T_{m n}$ containing $N_{m n}$ sampling points, so that the new sampling interval $N_t$ could be equal to the least common multiple of all $N_{m n}$ which could be determined in the time frequency spectrum. Since $t_0$ is the initial sampling points, the continuous target model expressed by equation $(34)$ can be rewritten as,
\begin{equation}
\begin{aligned}
\widetilde{\mathcal{T}}\left(x_{o}\right) &=\sum_{m=0}^{\infty} \mathcal{T}_{m}\left(x_{o}\right) \\
& \times \exp \left[j 2 k_{\lambda} W_{m}\left(x_{o}\right) \eta\left(t_{0}\right)\right]
\end{aligned}
\end{equation}
where, $\eta_{m}\left(t_{0}\right)$ is a non-time varying item. During all of the sampling points in the new time sequence, the phase plane remains unchanged which means 
\begin{equation}
\begin{aligned}
&\eta_{m}\left(t_{0}\right) = \eta_{m}\left(t_{0}+n N_t T_p\right)\\
&\widetilde{\mathcal{T}}\left(x_{o}, t_0\right) = \widetilde{\mathcal{T}}\left(x_{o}, t_0+n N_t T_p\right),n=0,1,2,...
\end{aligned}
\end{equation}where $T_p$ is the sampling interval between adjacent sampling points, $n$ is the serial number of new time sequence. Supposing that the output signal of the photoelectric receiver contains a total of $N_s$ sampling points, then $n=\left\lfloor\frac{N_{s}}{N_{t}}\right\rfloor$. Thus the first-order field correlation function in the continuous detecting scenarios can be rewritten as
\begin{equation}
\begin{aligned}
&G^{(1)}_{x_r}\left(x_{c}\right) \\
&\propto\left\langle\begin{array}{l}
\exp \left(-j k_{\lambda} Z_{2}\right) \int_{x_{s}^{\prime}} d x_{s}^{\prime} \int_{x_{s}} d x_{s} \int x_{x_{o}} d x_{o} \\
\times E_{s}\left(x_{s}\right) E_{s}^{\prime}\left(x_{s}\right) \sum_{m=0}^{\infty} \mathcal{T}_{m}\left(x_{o}\right)  \\
\times \exp \left[j 2 k_{\lambda} W_{m}\left(x_{o}\right) \eta\left(t_{0}\right)\right] \widetilde{U}\left(x_{r}, x_{c}\right) \\
\times \exp \left[\frac{j k_{\lambda}}{2 Z_{1}}\left(x_{o}^{2}-x_{c}^{2}\right)\right] \exp \left[\frac{j k_{\lambda}}{2 Z_{1}}\left(x_{c}-x_{o}\right) 2 x_{s}\right] \\
\times \exp \left[\frac{j k_{\lambda}}{2 Z_{2}}\left(x_{r}-x_{o}\right)^{2}\right]
\end{array}\right\rangle
\end{aligned}
\end{equation}
When the virtual modulation function is rewritten as
\begin{equation}
\widetilde{U}_{x_r}\left( x_{c}\right)=\exp \left(j k_{\lambda} Z_{2}\right) \exp \left[-\frac{j k_{\lambda}}{2 Z_{2}}\left(x_{r}-x_{c}\right)^{2}\right]
\end{equation}
the first-order field correlation between new sampled electronic signal $i_W(t)$ and its corresponding new reference field function can get the result of the equivalent complex valued target, which has illustrated in preceding section. As a result, function $(42)$ can be arranged as follows,
\begin{equation}
\begin{aligned}
&G^{(1)}\left(x_{c}\right) \\
&\propto \sum_{m=1}^{\infty} \mathcal{T}_m\left(x_{c}\right) \exp \left[j 2 k_{\lambda} W_{m}\left(x_{c}\right) \eta_{m}\left(t_{0}\right)\right]
\end{aligned}
\end{equation}
Equation $(44)$ and Fig. 9 show that there are $N_t - 1$ sequences could be obtained through the interval sampling method theoretically with the change of the initial sampling point $t_0$, it means that there are also $N_t-1$ final images can be obtained through the first-order field correlation. These images can demonstrate how the complex target plane changes over time $T_p N_t$, the sequences of images can also provides some target micro-vibration characteristics for further imaging processing.\\

Recalling the complex image derived from field correlation, the first-order field correlation result consists solely of $M$ principal modes and their corresponding reflective coefficient at time $t_0$. The $M$ principal modes have linear superposition in the same plane for this complex image, as demonstrated by the two-dimensional Fourier transform below
\begin{equation}
|\mathcal{T}(u, v)|=\int_{-\infty}^{+\infty} \int_{-\infty}^{+\infty} \widetilde{\mathcal{T}}\left(x_{o}, t_0\right) \exp[{-j 2 \pi(u x+v y)}]  d x d y
\end{equation}
The $M$ principal modes can be represented discretely in the $K$-Space after two-dimensional Fourier transform. As $t_0$ changes, the frequency peak in $K$-Space will not change except for the phase, thus the $K$-Space image of reconstructed target image shows the relation between the target type and principal modes.\\
\begin{figure*}[htbp]
\centering
\includegraphics[width=2\columnwidth,height=0.55\linewidth]{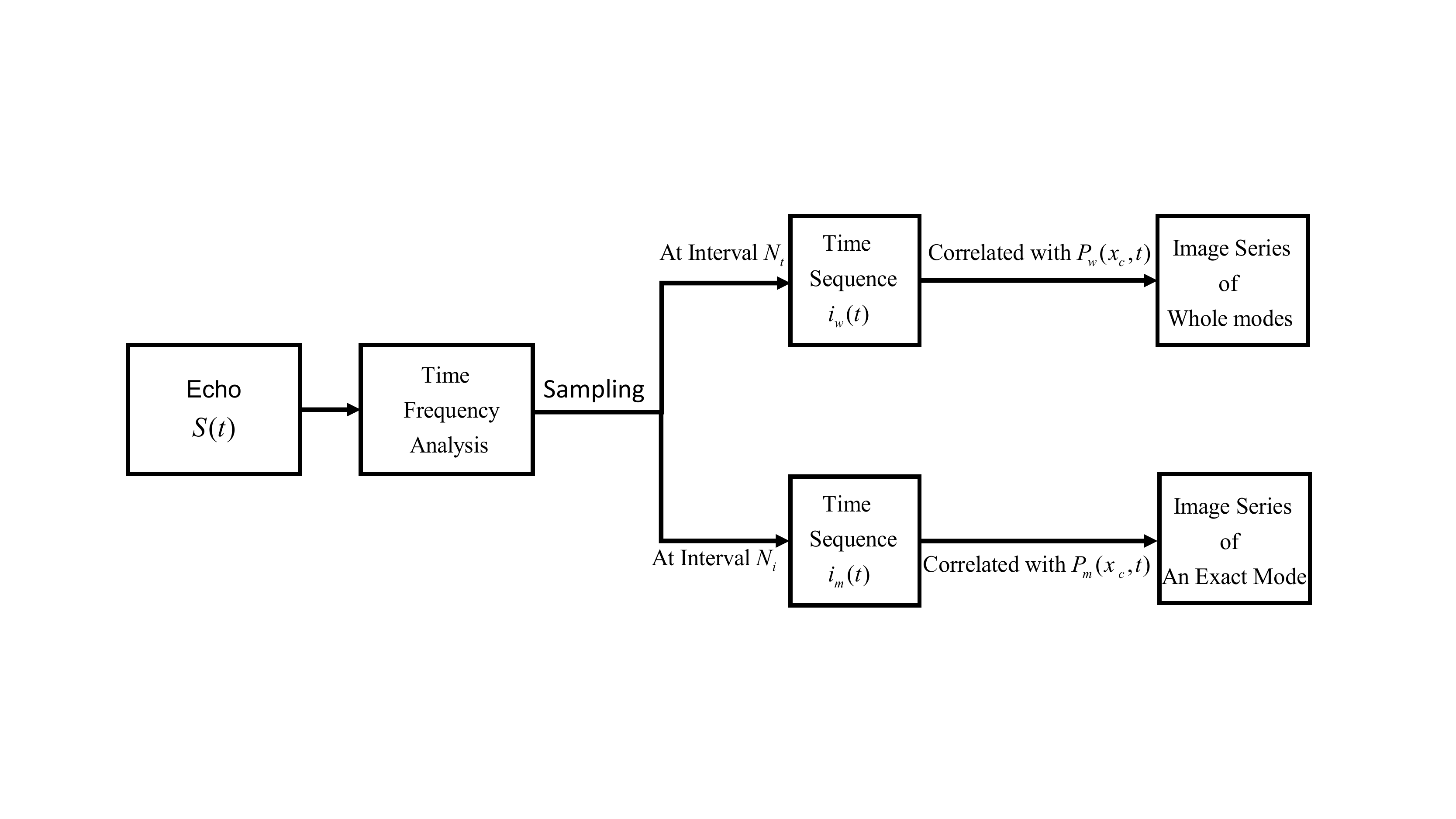}
\caption{The flow diagram about two different types of interval sampling method, the type one method sampled at the interval $N_t$ can finally obtain the linear superposition of all principal modes images at time $t_0$, and from $2-D$ Fourier transform, all of different principal modes have their corresponding independent frequency peak in $K$-space. And type two method shown in second row has new time interval $N_i$, the final images obtained by this way can give the spatial distribution of each frequency detected in the time frequency spectrum without spatial properties. }
\label{fig：1}
\end{figure*}\\

$\boldsymbol{Type 2}$: The method mentioned in $\boldsymbol{Type 1}$ whose new sampling interval $N_t$ is equal to the least common multiple of all $N_{m n}$, can fully reconstruct the superimposed image of all the principal modes with periodical changes. However, the time frequency characteristics have been smoothed due to the interval sampling, the spatial distribution of each frequency vibration cannot be shown in the results. The new sampling time interval can be set to equal to $T_{m n}$ to solve the space distribution of each single principal mode problem, where $T_{m n}$ is the period time corresponding to the $m$th principal mode and $n$-order frequency. Because the periodical time-varying function $\eta(t)$ can be expanded by the Fourier series, the fundamental mode period $T_m$ is integer multiples to all the high order vibration period $T_{m n}$. Periodical sampling of the fundamental mode vibration can also be carried out by periodical sampling of higher order mode vibrations. The same as $\boldsymbol{Type 1}$, the mode frequency $f_{m}$ can be determined in the time frequency spectrum. \\

Assuming that the frequency awaits to be solved is $f_{i}$, thus the new sampling time interval is $T_{i }$, the sampling points number containing in $T_{i}$  is $N_{i}$. According to corresponding relations shown in Fig. 10, the corresponding speckle field patterns sequence can be found for the new sampling sequence. Since $t_0$ is the initial sampling points, the continuous target model can be rewritten from equation $(34)$ as,
\begin{equation}
\begin{aligned}
\widetilde{\mathcal{T}}\left(x_{o},n\right) &=\mathcal{T}_{i}\left(x_{o}\right) \exp \left[j 2 k_{\lambda} W_{i}\left(x_{o}\right) \eta_i\left(t_{0}\right)\right] \\
&+\sum_{m=0}^{\infty} \mathcal{T}_{m}\left(x_{o}\right) \\
& \times \exp \left[j 2 k_{\lambda} W_{m}\left(x_{o}\right) \eta_m\left(t_{0}+n T_p N_i\right)\right], m \neq i
\end{aligned}
\end{equation}
where $n$ is a serial number of new sequence sampling points, $n=\left\lfloor\frac{N_{s}}{N_{i}}\right\rfloor$ in this scenario. The target model contains the time-invariant items of the principal mode to be measured and time-varying coupling items of other modes as shown in equation $(46)$. Substituting the result of equation $(46)$ into the first-order field correlation function
\begin{equation}
\begin{aligned}
&G^{(1)}_{x_r}\left(x_{c},n\right) \propto \mathcal{T}_{i}\left(x_{c}\right)\exp[j 2 k_{\lambda} W_i(x_c) \eta_i(t_0)]\\
&+\left\langle\begin{array}{l}
\exp \left(-j k_{\lambda} Z_{2}\right) \int_{x_{s}^{\prime}} d x_{s}^{\prime} \int_{x_{s}} d x_{s} \int x_{x_{o}} d x_{o} \\
\times E_{s}\left(x_{s}\right) E_{s}^{\prime}\left(x_{s}\right) \sum_{m=0}^{\infty} \mathcal{T}_{m}\left(x_{o}\right)  \\
\times \exp \left[j 2 k_{\lambda} W_{m}\left(x_{o}\right) \eta\left(t_{0} + n T_p N_i\right)\right] \widetilde{U}_{x_r}\left( x_{c}\right) \\
\times \exp \left[\frac{j k_{\lambda}}{2 Z_{1}}\left(x_{o}^{2}-x_{c}^{2}\right)\right] \exp \left[\frac{j k_{\lambda}}{2 Z_{1}}\left(x_{c}-x_{o}\right) 2 x_{s}\right] \\
\times \exp \left[\frac{j k_{\lambda}}{2 Z_{2}}\left(x_{r}-x_{o}\right)^{2}\right]
\end{array}\right\rangle
\end{aligned}
\end{equation}
where $m\neq i$. It can be seen that equation $(47)$ is consist of a constant complex item which is the space distribution of the filtering target principal mode $W_i(x_o)$ and a residual item which is from the other modes coherence superposition at the interval $T_{i }$. Since these principal modes have coherent aliasing in two dimensions. The same as discrete target reconstruction, there are other principal modes have projection residues on the principal mode $W_i(x_o)$ due to their non-orthogonality. Whereas in ideal infinite sampling ensemble average situation, these residual mathematical expected value is equal to zero. Consequently the field correlation function can be rewritten as
\begin{equation}
G^{(1)}\left(x_{c}\right) \propto \mathcal{T}_{i}\left(x_{c}\right)\exp[j 2 k_{\lambda} W_i(x_c) \eta_i(t_0)]
\end{equation}
\begin{figure*}[t]
\centering
\includegraphics[width=2\columnwidth,height=0.32\linewidth]{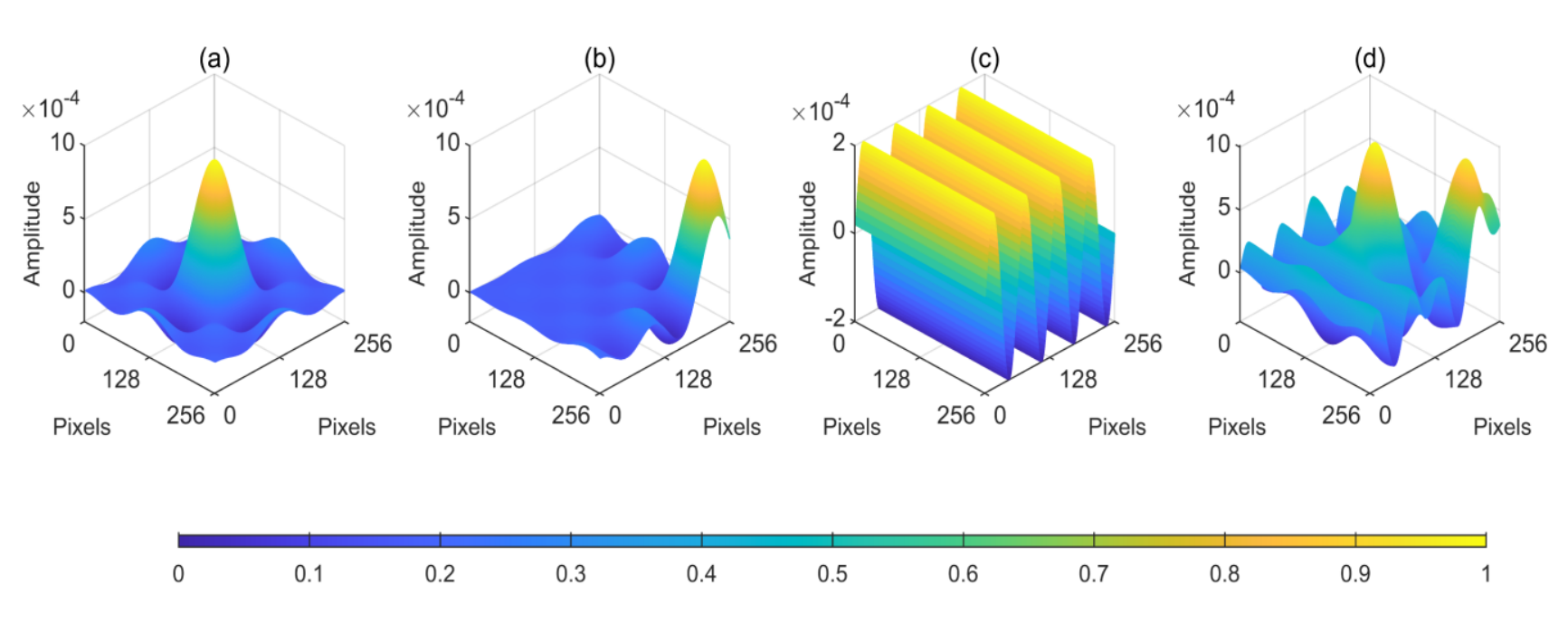}
\caption{The principal modes images at time $t_0$, modes $(a)$ to $(c)$ are the three different principal modes respectively and $(d)$ is the final image of these three modes superposition, where mode $(a)$ and mode $(b)$ is the same forced vibration, it can be seen as the target plane has two identical type of excitation sources in the different location. Mode $(c)$ is a type of free vibration, the vibration of this principal mode is not forced by external source, and there is usually a strong material correlation between free vibration and target characteristics. Since it is realized under the framework of optical detection system, the amplitude scale that can be detected is smaller than the wavelength of laser, which is 1550 nanometer in this simulation  }
\label{fig：1}
\end{figure*}\\

 \begin{figure*}[htbp]
\centering
\includegraphics[width=2\columnwidth,height=0.32\linewidth]{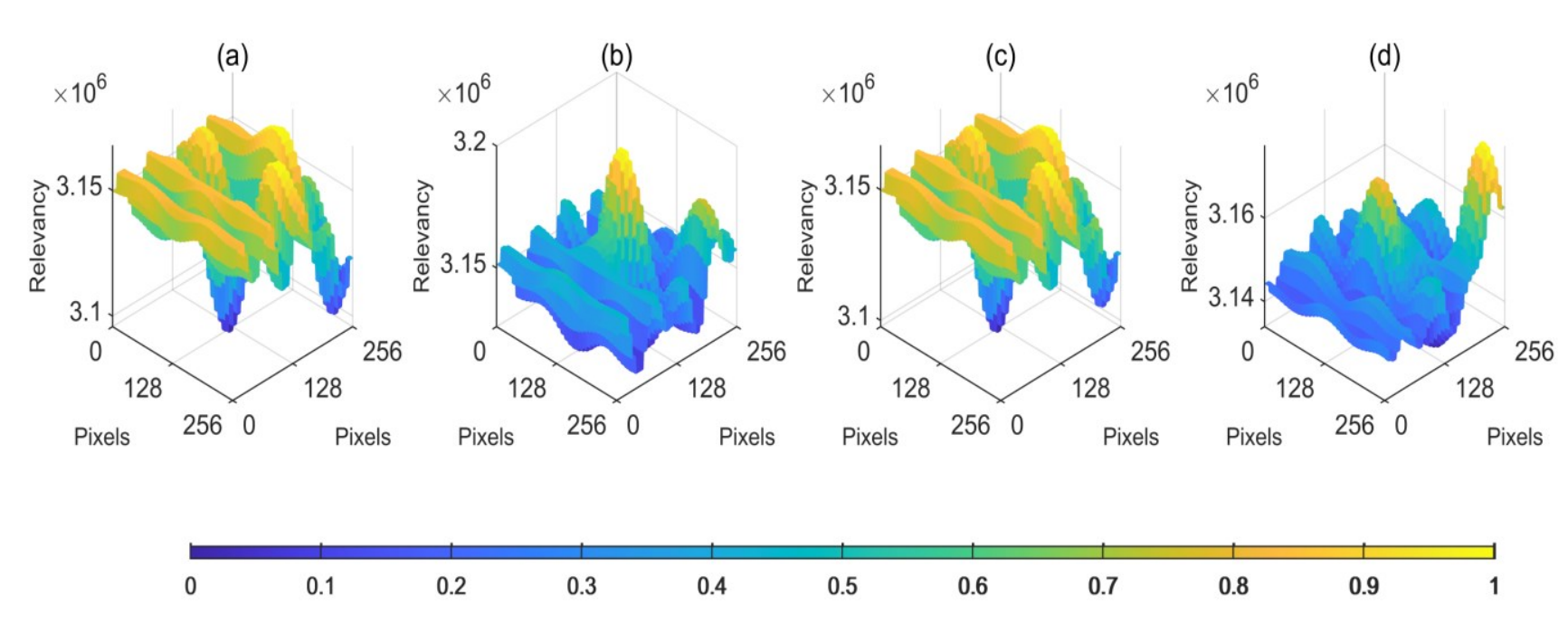}
\caption{The reconstruction results of principal modes at different time points , from $(a)$ to $(d)$ shows the final image of these three modes superposition changes during a single period, and if sampling rate is high enough, the step interval of $t_0$ can be shortened and a finer variation image of the target surface can be obtained.}
\label{fig：1}
\end{figure*}

In this method, every vibration space distributions can be shown separately in images, and the same as $\boldsymbol{Type 1}$ as shown in Fig. 9, $N_i -1$ time sequences can be acquired by selecting different initial sampling point $t_0$. A series of resulting images can shows the complex plane changes of a single principal mode $W_i(x_o)$ in a period $T_{i}$. The result of two-dimensional Fourier Transform is not accurate considering that the existence of the residual items in the resulting mode, thus the position in $K$-Space is inaccurate for the Fourier spectrum analysis. 

\subsection {Continuous model reconstructing simulation results}
 \begin{figure*}[htbp]
\centering
\includegraphics[width=2\columnwidth,height=0.9\linewidth]{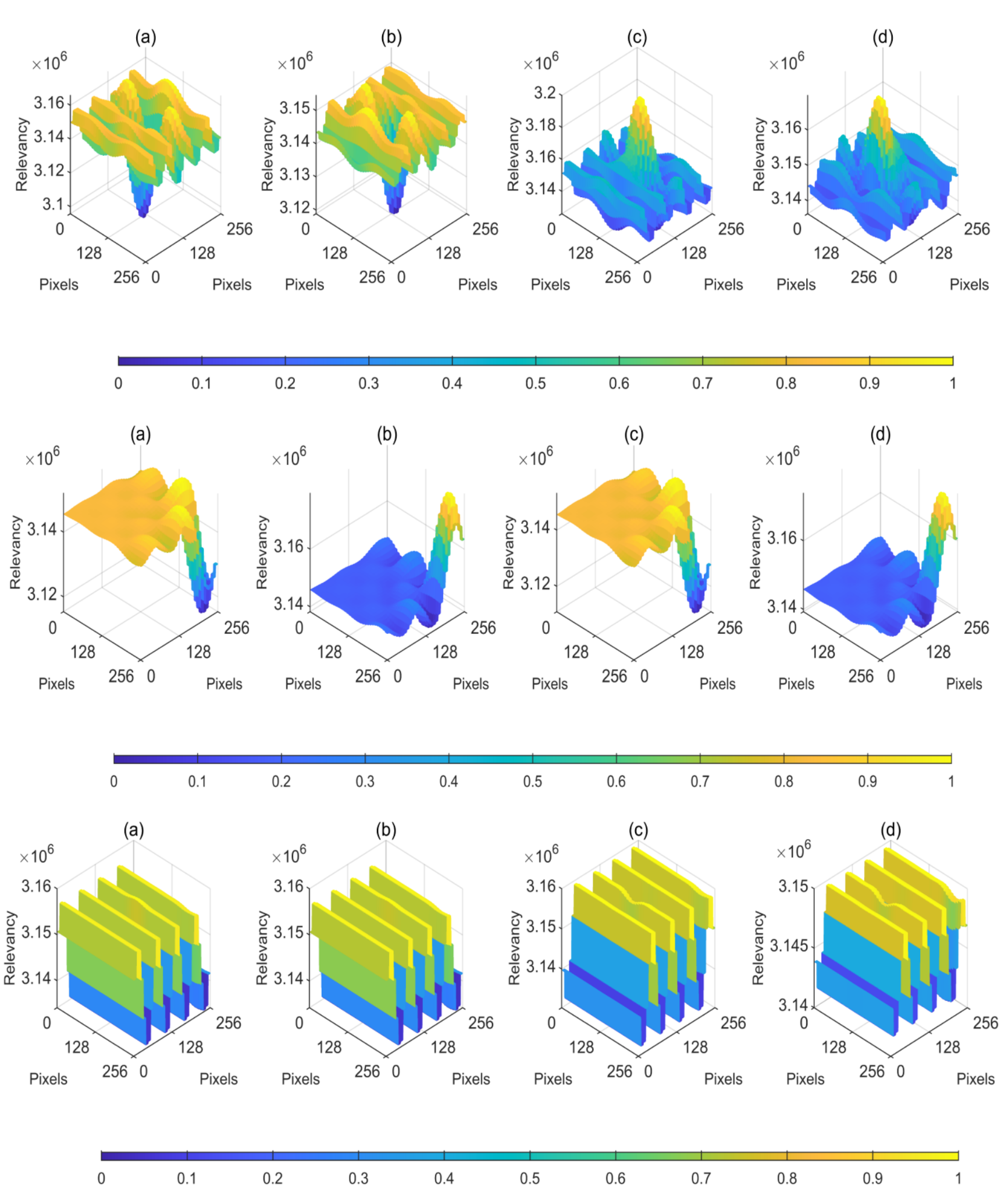}
\caption{This figure shows the complex plane of three principal modes change in a period, and different rows images are acquired from their own time sequences sampled at different sampling intervals. Because of the insufficient sampling rate of realistic systems, the residual projection from the other principal modes to the mode to be measured are reserved in the final images, set the first row images as an example, the free vibration distribution can be clearly seen and except for the center are of forced vibration, the transverse propagation of this forced vibration is submerged with the residual items. }
\label{fig：1}
\end{figure*}

The optical system is used as the sensor platform in continuous target detection scenario simulations. The configuration of the system is identical as the sensing systems in discrete target detecting scenarios except for the target model. As we mentioned before, the scale of the continuous targets is much greater than the single resolution of the speckle field. In order to have a better display of the simulation effect, the entire field of view has set to a reflection plane where the scattering points have the same reflectivity. The continuous target parameters are shown in Table.\uppercase\expandafter{\romannumeral3} 

\begin{table}[h]
\centering
\caption{PARAMETER SETTING FOR DIFFERENT VIBRATION MODES}
 \begin{tabular}{|c|c|c|c|} 
 \hline
 Parameters & Mode 1 & Mode 2 & Mode 3 \\ 
 \hline
 Principal Vibration&Forced & Forced & Free\\
 \hline
 Attenuation&Dirac&Dirac &Sinusoidal \\
 \hline
 
 Distribution&Center&Northeastern corner&Uniform\\
 \hline
 Amplitude(m) & 100e-9 & 500e-9 & 1200e-9 \\ 
 \hline
 Initial Phase(rad) & 0 & $\pi/4$ & $5\pi/6$ \\
 \hline
 Frequency(Hz) & 2 & 4 & 5 \\
 \hline
 \end{tabular}
 \label{table: 1}
\end{table}

As shown in Fig. 12, two types of principal modes, which are typically observed in practical circumstances, are put on the continuous target. Two same forced vibration source centers are placed in different positions to test the spatial resolving ability towards the same patterns, such forced vibration sources are common in realistic detecting scenarios, such as the vibration of an engine, the beating of the heart, and the swing of a mechanical arm or a human arm. In addition, another type of free vibration is placed on the target surface; this type of principal mode vibration is more impacted by material properties.\\

The following are the premises of this continuous target vibration detection simulation:
\begin{enumerate}
\item All of these modes are stationary wave vibration, traveling wave systems are not discussed in this study. 
\item The amplitude of the forced vibration is far greater than the free vibration normally, to make the final image more obvious, the times of amplitudes of two types vibration is limited within ten. 
\item The time-varying vibration function $\eta_m(t)$ corresponding to $m$th principal mode in this scene is set as a simple harmonic vibration for simplification.  
\end{enumerate}

The time frequency analysis towards the receiving signal from the time frequency spectrum is necessary for reconstructing the whole principal modes spatial distribution. According to the frequency obtained from time-frequency spectrum, the required time sampling interval $N_t$ and $N_i$ can be calculated. Since we propose two types of reconstruction methods for different requirements in continuous target detecting scenarios, the result can be divided into two parts.
$\boldsymbol{Type 1}$, we can see there are three vibrations motion detected in the field of view as shown in Fig. 13, which can be recorded as $f_m,m=1,2,3$. Combining with the sampling rate of the receiving system, the number of sampling points corresponding to each vibration period can be obtained, which can be recorded as $N_m, m=1,2,3$. According to the time interval sampling method introduced in Fig. 9, the least common multiple $N_t$ can be calculated from $N_m$. The new time sequence obtained by sampling at interval $N_t$ points can be first-order correlated with the reference radiation field corresponding to the sampling time points. In a periodical sampling points $N_t$, the change of the principal modes on the whole target plane can be obtained by stepping the position of the first sampling point.\\

The time varying in a period of three principal modes can be seen explicitly as shown in Fig. 13. Taking $t_0$ equal to four different time points at the same interval in a period. All of the images can theoretically be integrated into a video to better study the micro-vibrations characteristics of the target surface if the sampling number is large enough. Even though the image sequence of all principal modes can be clearly observed, there is still no one-to-one correspondence between frequencies measured from echo time frequency analysis and the spatial distributions of principal modes, but the method introduced in $\boldsymbol{Type 2}$ can solve this problem.
Because the three vibrations motion detected in time-frequency spectrum are recorded as $f_m$. Assuming $N_i=N_m$, $m=1,2,3$, for the new time sampling points interval $N_i$. For a certain frequency $f_i$, its corresponding spatial distribution can be obtained. All the three types of principal modes can be reconstructed as shown in first row of Fig. 14. Similarly, we set $t_0$ equal to four different time points at the same interval in a period, then three sequences of three principal modes are obtained. Because the time sampling points obtained are far less than the complete time sampling sequence, the new time sequence involved in the first-order field correlation falls far short of the requirement for an ensemble average, as shown in equation $(47)$, the residual item of the other modes cannot be completely eliminated. Thus in the non-ideal situations, equation $(47)$ can be rewritten as 
\begin{equation}
\begin{aligned}
&G^{(1)}\left(x_{c}\right) \propto \mathcal{T}_{i}\left(x_{c}\right) \exp \left[j 2 k_{\lambda} W_{i}\left(x_{c}\right) \eta_{i}\left(t_{0}\right)\right] \\
&+\frac{1}{n} \sum_{n=1}^{N} \sum_{m=0}^{\infty} \mathcal{T}_{m}\left(x_{c}\right) \exp \left[j 2 k_{\lambda} W_{m}\left(x_{c}\right) \eta\left(t_{0}+n T_{p} N_{i}\right)\right]
\end{aligned}
\end{equation}
Where $N$ is the final serial number of the new sampled time sequence. According to equation $(49)$ and  the first row of Fig. 14, the residual item caused by the free vibration principal mode can be observed clearly, except for the forced vibration center, other propagation characteristics of the target plane are obscured by the residuals of other principal modes. Meanwhile, in the third row of Fig. 14, the two forced vibration centers can also be barely observed in the images. \\
Generally speaking, the amplitude scale of the vibration usually reaches the micron level.The vibration of this magnitude level can be detected in the optical system, which is very important for future measurements of the optical system's material characteristics. For microwave systems, the detection of this free vibration mode is also useful in some scenarios, such as the oscillation of a sea-crossing bridge or a skyscraper.\\

% === FIG : SWEEP VG
%\begin{figure}[ht!]
%\centering
%\includegraphics[width=3.4in]{pdf/15.pdf}
%\caption{Measured conversion efficiency and drain DC voltage versus input power for several DC gate voltage biases. For this data, $R_{DC}=58\Omega$ and $Z_g(f_0)=\left(230+j10\right)\Omega$.}
%\label{VG_sweep_final}
%\end{figure}

\section{Conclusion}
In this work, the first-order field correlation imaging mode was first derived from a theoretical analysis of the target motion. We then proposed the micro-vibration mode reconstruction algorithm on the basis of the first-order field correlation. The feasibility of this theory is verified by numerical simulations. To the best of our knowledge, this is the first time for sensing systems to reconstruct the temporal-spatial distributions of  different vibrating target in a theoretical method. Simulation results showed that the target mode temporal-spatial distributions are well reconstructed. The proposed method in this study has a good combination of time frequency spectrum filtering and spatial imaging compared to conventional imaging systems. Hence this study provides the spatial distribution information and time domain information simultaneously. Due to the fact that we have only established the mathematical analytic model for the imaging process, there is still a great deal of work to be done on the effects, such as a more precise vibration mode, a method can overcome the residual on imaging caused by inadequate sampling, and a more efficient estimation method of target surface radiation field.

\section*{Appendix}
The ratio of transverse strain to longitudinal strain is known as the Poisson's ratio of solid materials according to the solid material general feature\cite{Contin1}. Generalizing the equation of wave motion to a two-dimensional plane when the turning radius is rectangular section
\begin{equation}
\nabla^{4} w+\frac{3 \rho\left(1-\mu^{2}\right)}{E h^{2}} \frac{\partial^{2} w}{\partial t^{2}}=0
\end{equation}
Where $\nabla^{4}$ is the quadruple-harmonic operator, $E$ is Young modulus of different materials, $\mu$ is its Poisson's ratio, $\rho$ is its intensity, $2h$ is the thickness, and $w$ represents the vibration function in the time-domain. Under the Kirchhoff $G$ model, the target can be considered as a sheet when its thickness is substantially smaller than its surface size. This part establishes the small deflection bending theory of elastic sheets and provides an accurate model of transverse vibration of rectangular plates. For other types of objects, the vibration model is adaptable by modifying the initial conditions and boundary conditions, which will not be classified or discussed in this study. The basic assumptions can be summarized as follow: the middle plane bisecting the plate thickness is the central plane, set central plane as $xOy$ plane, the $Rt$ coordinate space system is established as shown in Fig. 6. When a sheet bends, the central plane becomes a curved surface known as an elastic surface, and the displacement of a random point on this plane is $u$, $v$, and $w$, which correspond to the directions $x$, $y$, and $z$, respectively, where $w$ is the deflection.\\
\begin{enumerate}
\item Normal line hypothesis : The normal line which is perpendicular to the central plane remains straight in the whole procession, and also remains perpendicular to the curved surface of elasticity. In the other way, even though when the transverse shear stresses $\tau_{x z}$ and $\tau_{y z}$ are nonzero, the transverse shear deformation $\gamma_{x z}$ and $\gamma_{y z}$ are ignored.
\item The internal stresses of a bending plate are mainly $\sigma_{x}$, $\sigma_{y}$ and $\tau_{x y}$, and the secondary stresses are $\tau_{x z}$ and $\tau_{y z}$, the minimum stress is $\sigma_z$.
\item The change of thickness should be ignored, i.e. $\varepsilon_{z}=0$, so that every points on the line perpendicular to the elastic plate could have the same transverse  displacement $w$, which indicates that $w$ has no relation with $z$.
\item Deflection $w$ should be much less than the thickness of the sheet $h$, that is to say, all points on the surface of the central plane remain on the central plane during the target deformation.
\end{enumerate}
The rectangle infinitesimals $dxdy$ on the central plane can be replaced by the infinitesimal bodies $hdxdy$ based on the hypotheses listed above , so that the differential equation of the sheet's transverse vibration can be written as
\begin{equation}
D_{0} \nabla^{4} w+\rho h \frac{\partial^{2} w}{\partial t^{2}}=p(x, y, t)
\end{equation}
where $p(x,y,t)$ is a stress function, which represents the forced vibration on the target sheet, and $D_{0}=\frac{E h^{3}}{12\left(1-\mu^{2}\right)}$ is the flexural rigidity of the sheet, the equation should be transformed to a tensor form if the infinitesimal has not been replaced by rectangle infinitesimals,
\begin{equation}
\boldsymbol{\sigma}=\left(\begin{array}{l}
\sigma_{x} \\
\sigma_{y} \\
\tau_{x y}
\end{array}\right), \boldsymbol{D_1}=\left(\begin{array}{ccc}
1 & \mu & 0 \\
\mu & 1 & 0 \\
0 & 0 & \frac{1-\mu}{2}
\end{array}\right)
\end{equation}where $D_0$ represents scalar and $D_1$ represents non-scalar. Equation $(52)$ can be further organized as
\begin{equation}
\boldsymbol{\sigma}=\frac{E}{1-\mu^{2}} \boldsymbol{D_1} \varepsilon=\frac{E z}{1-\mu^{2}} \boldsymbol{D_1} \boldsymbol{\kappa}
\end{equation}where $\kappa$ is the transverse strain vector, thus the torque tensor can be determined as follow
\begin{equation}
\boldsymbol{M}=\frac{E}{1-\mu^{2}} \boldsymbol{D_1} \boldsymbol{\kappa} \int_{h / 2}^{-h / 2} z^{2} d z
\end{equation}Simplifying boundary conditions, the boundary of the sheet has no deflection and no bending moment $(x=0)$, thus we could assume that boundary conditions can be set as follow,
\begin{equation}
\left.w\right|_{x=0}=0,\left.\frac{\partial w}{\partial x}\right|_{x=0}=0
\end{equation}Assuming that all vibrations composed by the forced vibrations can be seen as a linear superposition of free vibrations, thus we can discuss the free vibration of sheets first, when $p(x, y, t)=0$, the free vibration function can be acquired
\begin{equation}
D_{0} \nabla^{4} w+\rho h \frac{\partial^{2} w}{\partial t^{2}}=0
\end{equation}
The vibration function should have a formal solution of separation of variables,
\begin{equation}
w(x, y, t)=W(x, y) \sin (\omega t+\phi)
\end{equation}
where $W(x,y)$ is the principal mode, so $\omega$ is the corresponding characteristic frequency. Subtitling equation $(58)$ into equation $(57)$,
\begin{equation}
\nabla^{4} W-\beta^{4} W=0
\end{equation}
where $\beta^{4}=\frac{\rho h}{D_{0}} \omega^{2}$, and because of the homogeneity of the boundary conditions for $W$, $\omega$ can be rewritten as $W$ directly. For the case of four-sided supporting, the exact principal mode vibration expression can be obtained as ,
\begin{equation}
W_{i, j}(x, y)=A_{i, j} \sin \frac{i \pi x}{a} \sin \frac{j \pi y}{b}, i, j=1,2,3, \ldots
\end{equation}
where $A_{i j}$ is the amplitude corresponding to different order vibration. From equation $(59)$ we can know that if we need the final form of principal mode, we need to solve the differential equation $(58)$, and $\beta$ should satisfy the boundary condition
\begin{equation}
\left[\left(\frac{i \pi}{a}\right)^{2}+\left(\frac{j \pi}{b}\right)^{2}\right]^{2}-\beta^{4}=0
\end{equation}
which can be rewritten as the form of eigenfrequency,
\begin{equation}
\omega_{i, j}=\pi^{2}\left(\frac{i^{2}}{a^{2}}+\frac{j^{2}}{b^{2}}\right)
\end{equation}
The precise orthogonality of eigenfrequency among the primary vibration can guarantee the decomposition is regular. However, our method of detection focus on the whole plane, we only need the variations form of the differential equation on the whole sheet plane$(\Omega)$ :
\begin{equation}
\iint_{\Omega}\left(D_{0} \nabla^{4} W-\omega^{2} \rho h W\right) \delta W d x d y=0
\end{equation}The functional characteristic values of plate vibrations can also be obtained:
\begin{equation}
\omega^{2}=\frac{\iint_{\Omega} D_{0} \boldsymbol{\kappa}^{T} \boldsymbol{D_1} \boldsymbol{\kappa} d x d y}{\iint_{\Omega} \rho h W^{2} d x d y}
\end{equation}
The principal modes $W_i(x,y)$ and $W_j(x,y)$ have two different natural frequencies $\omega_i$ and $\omega_j$ under the condition of linear indexes, thus $W$ can be written as
\begin{equation}
W=a_{i} W_{i}+a_{j} W_{j}
\end{equation}
where $a_i$ and $a_j$ are the constant coefficients of each components respectively, considering $\kappa$ is a linear tensor, the relationship between $W_i$, $W_j$ and $\kappa$ can be written as,
\begin{equation}
\boldsymbol{\kappa}(W)=a_{i} \boldsymbol{\kappa}\left(W_{i}\right)+a_{j} \boldsymbol{\kappa}\left(W_{j}\right)
\end{equation}
Substituting equations $(64)$ and $(65)$ into $(63)$, we can obtain
\begin{equation}
\begin{aligned}
&\iint_{\Omega} D_{0} \boldsymbol{\kappa}^{T} \boldsymbol{D_1} \boldsymbol{\kappa} d x d y \\
&=\iint_{\Omega} D_{0}\left[a_{i} \boldsymbol{\kappa}\left(W_{i}\right)+a_{j} \boldsymbol{\kappa}\left(W_{j}\right)\right]^{T} \\
&\quad \times \boldsymbol{D_1}\left[a_{i} \boldsymbol{\kappa}\left(W_{i}\right)+a_{j} \boldsymbol{\kappa}\left(W_{j}\right)\right] d x d y \\
&=a_{i}^{2} k_{i i}+a_{i} a_{j} k_{i j}+a_{j} a_{i} k_{j i}+a_{j}^{2} k_{j j}
\end{aligned}
\end{equation}and
\begin{equation}
\begin{aligned}
&\iint_{\Omega} \rho h W^{2} d x d y \\
&=\iint_{\Omega} \rho h\left(a_{i} W_{i}+a_{j} W_{j}\right)\left(a_{i} W_{i}+a_{j} W_{j}\right) d x d y \\
&=a_{i}^{2} m_{i i}+a_{i} a_{j} m_{i j}+a_{j} a_{i} m_{j i}+a_{j}^{2} m_{j j}
\end{aligned}
\end{equation}
The constants in equations $(66)$ and $(67)$ satisfy,
\begin{equation}
\begin{aligned}
&k_{m n}=k_{n m}=\iint_{\Omega} D_{0} \boldsymbol{\kappa}^{T}\left(W_{m}\right) \boldsymbol{D_1} \boldsymbol{\kappa}\left(W_{n}\right) d x d y \\
&m_{m n}=m_{n m}=\iint_{\Omega} \rho h W_{m} W_{n} d x d y
\end{aligned}
\end{equation}If the matrices $\boldsymbol{K}$,$\boldsymbol{M}$ and the vector $\boldsymbol{a}$ are
\begin{equation}
\boldsymbol{K}=\left[\begin{array}{ll}
k_{i i} & k_{i j} \\
k_{j i} & k_{j j}
\end{array}\right], \boldsymbol{M}=\left[\begin{array}{ll}
m_{i i} & m_{i j} \\
m_{j i} & m_{j j}
\end{array}\right], \boldsymbol{a}=\left[\begin{array}{l}
a_{i} \\
a_{j}
\end{array}\right]
\end{equation}
the characteristic value function $(63)$ can be conveniently given by the quadratic form of the vector,
\begin{equation}
\omega^{2}=\frac{\boldsymbol{a}^{T} \boldsymbol{K}\boldsymbol{a}}{\boldsymbol{a}^{T} \boldsymbol{M} \boldsymbol{a}}
\end{equation}
The vector $\boldsymbol{a}$ can be seen as a variable in equation $(70)$, when the derivative of the characteristic value function is zero
\begin{equation}
\delta\left(\omega^{2}\right)=0
\end{equation}
then we can have
\begin{equation}
\frac{1}{\left(\boldsymbol{a}^{T} \boldsymbol{M} \boldsymbol{a}\right)^{2}}\left[\left(\boldsymbol{a}^{T} \boldsymbol{M}  \boldsymbol{a}\right) \delta\left(\boldsymbol{a}^{T} \boldsymbol{K}  \boldsymbol{a}\right)-\left(\boldsymbol{a}^{T} \boldsymbol{K}  \boldsymbol{a}\right) \delta\left(\boldsymbol{a}^{T} \boldsymbol{M}  \boldsymbol{a}\right)\right]=0
\end{equation}
Consequently, equation $(72)$ can be simplified on the based of equation $(70)$
\begin{equation}
\delta^{T}(\boldsymbol{a})\left(\boldsymbol{K}-\boldsymbol{\omega}^{2}\boldsymbol{M} \right) \boldsymbol{a}=0
\end{equation}
According to the arbitrary of $\delta^{T}(\alpha)$, 
\begin{equation}
\left(\boldsymbol{K}-\boldsymbol{\omega}^{2}\boldsymbol{M} \right) \boldsymbol{a}=0
\end{equation}
Assigning $a_i=0,a_j=1$ and $a_j=0,a_i=1$ respectively, the difference is 
\begin{equation}
\left(\omega_{i}^{2}-\omega_{j}^{2}\right) m_{i j}=0
\end{equation}
if $\omega_{i}^{2} \neq \omega_{j}^{2}$, then $m_{i j}=\delta_{i j}$. For the same, $k_{i j}=\delta_{i j}$. When the characteristic value equation gives multiple roots, we can also make the $n$th multiple roots give $n$th orthogonal principal modes through reasonable orthogonalization.  Thus the orthogonality among cross-indexes principal modes can be given as
\begin{equation}
\begin{aligned}
&\iint_{\Omega} D_{0} \kappa^{T}\left(W_{i j}\right) D_{1} \kappa\left(W_{m n}\right) d x d y=\omega_{m n}^{2} \delta_{i m} \delta_{j n} \\
&\iint_{\Omega} \rho h W_{i j} W_{m n} d x d y=\delta_{i m} \delta_{j n}
\end{aligned}
\end{equation}
The forced vibration of rectangular sheet can be determined by the oscillator composition after we have the orthogonality among the principal modes on the surface of rectangular sheet. Expanding $w$ to bi-series as orthogonal principal modes
\begin{equation}
w(x, y, t)=\sum_{i=1}^{\infty} \sum_{j=1}^{\infty} W_{i j}(x, y) \eta_{i j}(t)
\end{equation}
Subtitling equation $(51)$,
\begin{equation}
\begin{aligned}
&D_{0} \sum_{i=1}^{\infty} \sum_{j=1}^{\infty}\left[\nabla^{4} W_{i j}(x, y)\right] \eta_{i j}(t) \\ 
& +\rho h \sum_{i=1}^{\infty} \sum_{j=1}^{\infty} W_{i j}(x, y) \ddot{\eta}_{i j}(t)=p(x, y, t)
\end{aligned}
\end{equation}
Multiplying with $W_{m n}(x,y)$ on both sides of equation $(78)$ and integrating $x$ and $y$ over $\Omega$,
\begin{equation}
\begin{aligned}
& \sum_{i=1}^{\infty} \sum_{j=1}^{\infty} \eta_{i j}(t) \iint_{\Omega} D_{0}\left[\nabla^{4} W_{i j}(x, y)\right] W_{m n}(x, y) \mathrm{d} x \mathrm{~d} y \\
& +\sum_{i=1}^{\infty} \sum_{j=1}^{\infty} \ddot{\eta}_{i j}(t) \iint_{\Omega} \rho h W_{i j}(x, y) W_{m n}(x, y) \mathrm{d} x \mathrm{~d} y \\
&= \iint_{\Omega} p(x, y, t) W_{m n}(x, y) \mathrm{d} x \mathrm{~d} y
\end{aligned}
\end{equation}
Then the canonical equation can be acquired by using conditions of orthogonality,
\begin{equation}
\begin{aligned}
& \ddot{\eta}_{m n}(t)+\omega_{m n}^{2} \eta_{m n}(t)=q_{m n}(t) \\ 
 \quad & q_{m n}(t)=\iint_{\Omega} p(x, y, t) W_{m n}(x, y) \mathrm{d} x \mathrm{~d} y
\end{aligned}
\end{equation}
When the sheet satisfy the initial condition
\begin{equation}
w(x, y, 0)=\lambda_{1}(x, y),\left.\quad \frac{\partial w}{\partial t}\right|_{t=0}=\lambda_{2}(x, y)
\end{equation}
then we can determine the initial condition in canonical coordinate
\begin{equation}
\begin{aligned}
&\eta_{m n}(0)=\iint_{\Omega} \rho h \lambda_{1}(x, y) W_{m n}(x, y) \mathrm{d} x \mathrm{~d} y \\
&\dot{\eta}_{m n}(0)=\iint \rho h \lambda_{2}(x, y) W_{m n}(x, y) \mathrm{d} x \mathrm{~d} y
\end{aligned}
\end{equation}
The solution of equation $(80)$ should be
\begin{equation}
\begin{aligned}
\eta_{m n}(t)&= \eta_{m n}(0) \cos \omega_{m n} t+\frac{\dot{\eta}_{m n}(0)}{\omega_{m n}} \sin \omega_{m n} t \\
&+\frac{1}{\omega_{m n}} \int_{0}^{t} q_{m n}(\tau) \sin \omega_{m n}(t-\tau) \mathrm{d} \tau
\end{aligned}
\end{equation}

% or


% ============================================
%\appendices
%Appendix one text goes here %\cite{Roberg2010}.

%\section{}

%\section*{Acknowledgment}

%The authors would like to thank D. Root for the loan of the SWAP. The SWAP that can ONLY be usefull in Boulder...

\ifCLASSOPTIONcaptionsoff
  \newpage
\fi

\vfill

\end{document}